\documentclass[acmtog]{acmart}
\acmSubmissionID{132}
\setcopyright{none}

\setcitestyle{square}

\usepackage{amsmath,amsthm,amssymb}
\usepackage{graphicx}
\usepackage{wrapfig}
\usepackage{subcaption}
\usepackage{color}
\usepackage{xspace}
\usepackage{overpic}
\usepackage{wrapfig}
\usepackage{enumitem}
\usepackage[ruled,vlined,linesnumbered]{algorithm2e}
\usepackage[title]{appendix}

\definecolor{turquoise}{cmyk}{0.65,0,0.1,0.1}
\definecolor{purple}{rgb}{0.65,0,0.65}
\definecolor{dark_green}{rgb}{0, 0.5, 0}
\definecolor{orange}{rgb}{0.8, 0.6, 0.2}
\definecolor{red}{rgb}{0.8, 0.2, 0.2}
\definecolor{brown}{rgb}{0.5, 0.16, 0.16}
\newcommand{\cy}[1]{{\color{black}#1}}
\newcommand{\kx}[1]{{\color{black}#1}}
\newcommand{\rz}[1]{{\color{black}#1}}
\newcommand{\sid}[1]{{\color{black}#1}}

\DeclareMathOperator*{\argmin}{argmin}

\newcommand{\SCORES}{\mbox{SCORES}\xspace}

\newcommand{\cB}{\mathcal{B}}

\newcommand{\cE}{\mathcal{E}}

\newcommand{\cH}{\mathcal{H}}

\graphicspath{{figures/img/}}

\newcommand{\mypara}{\paragraph}

\begin{document}

\title{\SCORES: Shape Composition with Recursive Substructure Priors}

\author{Chenyang Zhu}
\affiliation{%
  \institution{Simon Fraser University}
}
\affiliation{%
	\institution{National University of Defense Technology}
}
\author{Kai Xu}
\authornote{Corresponding author: kevin.kai.xu@gmail.com}
\affiliation{%
	\institution{National University of Defense Technology and Princeton University}
}
\author{Siddhartha Chaudhuri}
\affiliation{%
	\institution{Adobe Research}
}
\affiliation{%
	\institution{IIT Bombay}
}
\author{Renjiao Yi}
\affiliation{%
	\institution{Simon Fraser University}
}
\affiliation{%
	\institution{National University of Defense Technology}
}
\author{Hao Zhang}
\affiliation{%
	\institution{Simon Fraser University}
}

\renewcommand\shortauthors{C. Zhu et al}

\begin{abstract}
We introduce \SCORES, a {\em recursive neural network} for {\em shape composition}. Our network takes as input sets of parts from two or more source 3D shapes and a rough initial placement of the parts. It outputs an optimized part structure for the composed shape, leading to high-quality geometry construction. A unique feature of our composition network is that it is not merely learning how to connect parts. Our goal is to produce a coherent and {\em plausible} 3D shape, despite large incompatibilities among the input parts. The network may significantly alter the geometry and structure of the input parts and {\em synthesize} a novel shape structure based on the inputs, while adding or removing parts to minimize a structure plausibility loss. We design \SCORES as a {\em recursive autoencoder\/} network. During encoding, the input parts are recursively grouped to generate a root code. During synthesis, the root code is decoded, recursively, to produce a new, coherent part assembly. Assembled shape structures may be novel, with little global resemblance to training exemplars, yet have plausible substructures. \SCORES therefore learns a hierarchical {\em substructure shape prior\/} based on per-node losses. It is trained on structured shapes from ShapeNet, and is applied iteratively to reduce the plausibility loss. We show results of shape composition from multiple sources over different categories of man-made shapes and compare with state-of-the-art alternatives, demonstrating that our network can significantly expand the range of composable shapes for assembly-based modeling.

\end{abstract}

%
%







\begin{teaserfigure}
   \centering\includegraphics[width=0.99\textwidth,tics=100]{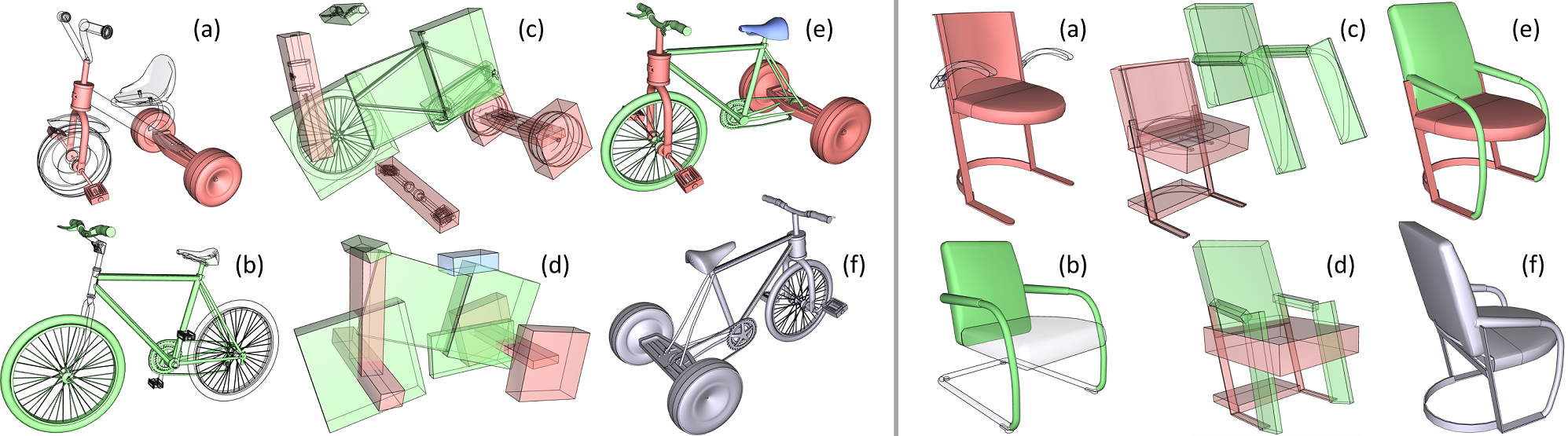}
   \caption{We introduce \SCORES, a neural network which learns structure fusion for 3D shape composition. \SCORES takes {\em box abstractions\/} of two sets of parts (red and green) from two source shapes (a-b), and a rough initial placement of the boxes (c), and outputs an optimized box structure (d), leading to quality geometry construction; see (e)-(f) for two views.
A unique feature of \SCORES is that it is not merely learning how to connect parts; the goal is to produce a plausible and coherent final shape structure, which may necessitate adding new parts (blue bicycle seat) or removing duplicates (red chair back). To handle creatively composed shapes, \SCORES learns a plausibility prior over {\em substructures} at various levels of abstraction, rather than complete shapes alone.
}
   \label{fig:teaser}
   \vspace{3mm}
\end{teaserfigure}

\setcopyright{acmcopyright}
\acmJournal{TOG}
\acmYear{2018}\acmVolume{37}\acmNumber{6}\acmArticle{1}\acmMonth{11} \acmDOI{10.1145/3272127.3275008}

\maketitle




\section{Introduction}
\label{sec:intro}

\begin{figure*}[t!] \centering
   \begin{overpic}[width=0.95\textwidth,tics=100]{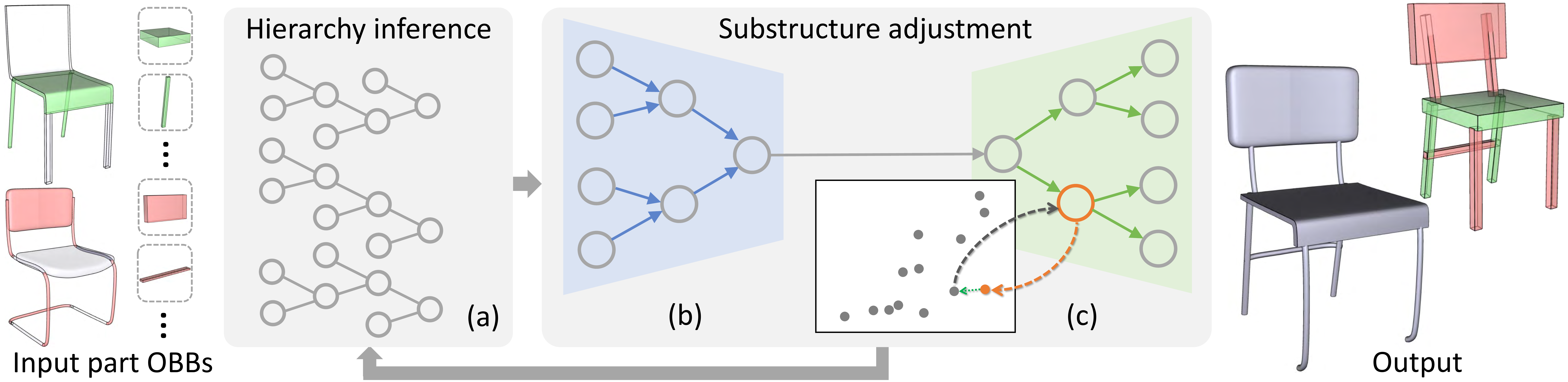}
   \end{overpic}
   \caption{\sid{Overview of our \SCORES shape composition pipeline, iterating over two key stages: (a) hierarchy inference, and (b-c) substructure adjustment. Adjustment comprises two passes: substructure embedding via bottom-up structure encoding (blue arrows); and substructure adjustment via top-down structure decoding (green arrows) and code adjustment (dashed arrows).
   The input consists of a group of parts, along with their OBBs, from two source shapes. Note that neither semantic nor relational information about the input parts is required throughout the pipeline; only that the source shapes are segmented.}}
   \label{fig:overview}
   \vspace{-2mm}
\end{figure*}

Composition-based shape synthesis has been one of the most frequently adopted modeling paradigms for virtual 3D shapes since the seminal work of Funkhouser et al.~\shortcite{funkhouser2004} on ``modeling by example''. Instead of crafting shape geometry from scratch with low-level curve and surface primitives, shape composition is a data-driven approach which {\em mixes-n-matches object parts} already available in a shape collection. Such an approach allows even novice users to model 3D shapes with detailed geometry and complex structure, and facilitates design space exploration~\cite{chaudhuri2010,xu2012,cohenor2016,ritchie2018}.

A primary challenge in mix-n-match modeling is how to resolve {\em incompatibilities} between selected parts to compose a well-structured, plausible 3D shape. Topological mismatches and significant positional misalignments between parts commonly cause such incompatibilities. From a modeling perspective, however, one often wants to combine incompatible parts in creative, yet sensible, ways to generate surprising and novel outcomes; see Figure~\ref{fig:teaser}.

Existing works on mix-n-match  modeling~\cite{chaudhuri2011,jain2012,kalogerakis2012,xu2012,laga2013,zheng2013} have mainly focused on discovering and retrieving parts which are geometrically, semantically, or functionally compatible with their counterparts. Considerably less effort has been devoted to address the {\em composition challenge} of
rearranging and deforming parts to resolve large discrepancies.


To address this challenge, we make the key observation that the ultimate goal of shape composition is to produce a final 3D shape that is {\em plausible\/} under a {\em structure prior}. While preserving the geometry and structure of the two (or more) sets of input parts is still an important criterion, the composition process should be allowed to significantly alter both attributes, possibly producing a novel overall structure for the composite shape. \rz{In particular, in light of possible large incompatibilities between the input parts, shape composition should not only be about connecting parts, it also has a {\em structure synthesis} aspect to it. The predominantly compositive process should also be able to {\em add\/} or {\em remove\/} parts to achieve the ultimate goal of mix-n-match shape modeling: to produce a plausible, coherent, and clean final 3D shape; see Figure~\ref{fig:teaser}.}
%

In this paper, we introduce a {\em machine learning\/} approach to shape composition. With a structure prior that is learned from training data, we aim to overcome limitations of existing, heuristic-driven part connection and substitution schemes, and expand the range of composable shape structures. The core of our learning approach is a novel {\em recursive neural network} or RvNN. RvNNs can be trained to learn parse trees for sentences or images for classification tasks~\cite{socher2011}. More recently, they have been used as generative models of global shape structures~\cite{li2017}. However, such a global prior is insufficient for shape composition. The new challenge is that the composed shapes may possess {\em novel structures\/} which do not globally match any exemplar.

To address this new challenge, we make another key observation about the {\em recurrent\/} nature of {\em substructures\/}~\cite{zheng2014}: salient part groups that frequently appear across a variety shapes and object categories. For instance, chairs, tables, and beds have different global structures but all contain a flat surface supported by legs, and bicycles and tricycles both combine a handlebar above a single front wheel. Even if a novel composite structure does not match any examplars globally, it often preserves recurrent substructures. Hence, our recursive neural architecture, called \SCORES, learns to explicitly model substructures at various levels of abstraction, rather than concentrating on a monolithic global prior. \rz{When composing shapes, we are opting to perform {\em substructure adjustments\/} since it is infeasible to rely on global structure optimization when composing incompatible structures. Thus we need to learn a manifold of valid substructures, where the substructure adjustment can be interpreted as ``docking'' onto the learned manifold.}

The input to our method is a set of parts from two or more source shapes, possibly belonging to different object categories. The output is a new composite shape, possibly with a novel overall structure. The input parts are only roughly aligned, and may have missing or redundant parts (Figure \ref{fig:teaser}). \SCORES formulates shape composition as a {\em hierarchical substructure optimization\/} problem, where the objective is to adjust part positions and geometry, and if necessary synthesize or remove parts, to account for the plausibility of each nested substructure rather than the global shape alone. Since the initial part alignment can be highly unreliable, we apply the trained network {\em iteratively} to progressively reduce the plausibility loss, interleaving hierarchy inference and substructure adjustment. To account for possible missing or redundant parts in the input, \SCORES can synthesize novel shape structures by adding or removing parts if the plausibility loss resulting from the current parts is too large to overcome by geometric adjustments alone.

The performance of our trained network is benchmarked on the ComplementMe dataset~\cite{sung2017}, a subset of the ShapeNet repository~\cite{chang2015_shapeNet}. The dataset contains more than 2000 3D shapes equipped with meaningful part segmentations for assembly-based shape modeling. Note that while the parts in ComplementMe are labeled, we {\em do not use the labels} in our learning and inference pipelines. We sample subsets of parts from dataset shapes, and merge them to form training and testing data. Performance is evaluated both with a structural plausibility measure based on projective shape comparison~\cite{zhu2017}, and by user studies.

In summary, our work makes the following contributions:
\vspace{-3pt}
\begin{itemize}
  \item The first machine learning approach to 3D shape composition, where a trained neural network is able to resolve significant geometric and topological incompatibilities between parts, even allowing part insertion and removal to maximize plausibility of the final composite shape,
  \item A formulation of shape composition with a plausibility objective based on substructure rectification,
  \item A novel RvNN architecture trained to learn a substructure prior and to minimize the plausibility loss via iterative hierarchy inference and substructure adjustment, and
  \item A benchmark for evaluating shape composition tasks.

\end{itemize}

\section{Related work}
\label{sec:related}


There is a significant body of work on 3D modeling via part-based shape composition, also known as assembly-based modeling. The approach has its roots in real-world design prototyping by piecing together existing building blocks, as exemplified in \mbox{``Kitbashing''} \cite{kitbashing:wiki}, or even children's mix-n-match flipbooks for creating fantastical creatures~\cite{ball1985}. Of course, the practice of constructing fixed designs with pre-fabricated components is even more widespread, e.g., in modular homes or IKEA furniture. Whether the goal is creative design, manufacturing ease, or transportation efficiency, a common thread in all forms of modular construction is the symbiosis of reuse and accessibility.
These advantages carry over to virtual shape modeling as well.

In this section, we discuss related works on assembly-based modeling. While most works have studied part {\em suggestions} for interactive modeling~\cite{chaudhuri2010,chaudhuri2011,sung2017}, 
we largely omit them since suggestions are not directly relevant to our work. However, we will discuss other aspects of these systems as appropriate.

\vspace{-1mm}
\mypara{Part connection.}
Several authors have studied how to seamlessly connect pairs of parts, assuming that the parts to be connected have already been selected and placed in close alignment. The main challenge is to deform and seal adjacent boundary loops of the parts to create smooth joins~\cite{funkhouser2004,sharf2006}.
The work of 
Takayama et al.~\shortcite{takayama2011} allow regions from one mesh to be grafted onto another {\em without} a pre-defined boundary loop on the target surface. Recent work by Duncan et al.~\shortcite{duncan2016} optimizes a shape collection for seamless part interchangeability.

The above methods use local geometric deformations to splice parts. A complementary line of work optimizes shape structure for better connectivity. Representative approaches set up an optimization problem to minimize the separations of likely connections between parts, e.g.~\cite{kalogerakis2012,schulz2014}.

In contrast to these works, our method aims to optimize the overall structure of a crudely assembled shape, both in geometry and in topology, to maximize a data-driven {\em plausibility} prior, optionally adding or removing parts for greater coherence and realism.


\vspace{-1mm}
\mypara{Part substitution and crossover.} 
The Modeling by Example system of Funkhouser et al.~\shortcite{funkhouser2004} enables users to retrieve parts by drawing simple proxies augmenting partially modeled shapes. Kraevoy et al.~\shortcite{kraevoy2007} automatically detect regions of exemplars matching a query shape to enable part exchange. Jain et al.~\shortcite{jain2012} and Alhashim et al.~\shortcite{alhashim2014} further automate the process by inferring a sequence of part substitutions to blend one shape into another. The Fit-and-Diverse system of Xu et al.~\shortcite{xu2012} evolves a population of shapes by exchanging parts between them.
Ritchie et al.~\shortcite{ritchie2018} learn a probabilistic program from the consistent hierarchies of exemplar shapes to generate new shapes with plausible topology and geometry. All of these systems require known correspondences between parts in different shapes. \rz{However, finding correspondences between topologically disparate shapes is challenging, since semantically correct correspondences may not even exist.}
In contrast, \SCORES is {\em correspondence-free}, does not require access to the complete source shapes, and can modify the structure to improve plausibility.

Zheng et al.~\shortcite{zheng2013} detect a small number of {\em manually-defined\/} substructure templates (e.g. a particular form of symmetric support) in a shape collection and permit part substitutions if the source and target parts conform to the same template. The employment of substructure modeling allows their work, as well as others, e.g.,~\cite{bokeloh2010}, to perform cross-category part substitution, like \SCORES. However, such substitutions were designed to preserve the global structure of the target shape. In contrast, \SCORES is not confined to (global) structure preservation. Moreover, it discovers and models a much larger variety of nested substructures directly from data, without human intervention, and learns plausible shape composition using a neural network.

%
%

\vspace{-1mm}
\mypara{Part placement.} Inferring the correct placement of a part newly added to an assembly is a challenging problem. Works such as that of Kalogerakis et al.~\shortcite{kalogerakis2012} address this with semantic annotations (e.g., ``wing attaches to body''). \SCORES relaxes this requirement by not needing such relation annotations at all: in fact, our method does not require any part labels. A very recent relevant work is the ComplementMe system of Sung et al.~\shortcite{sung2017}, which trains a ``placement network'' to predict where a newly suggested part should be placed in the assembly. Sung et al. are arguably the first to apply neural networks to part layout during shape composition. However, it is designed only to place a single part, and does not generalize to novel shapes that do not globally resemble any training exemplar.

\vspace{-1mm}
\mypara{\SCORES vs.~GRASS.} Our neural network, \SCORES, was inspired by GRASS~\cite{li2017}. However, the two RvNNs tackle different problems. GRASS is designed to synthesize 3D shapes globally resembling exemplars. \cy{It works ``intra-shape'', where symmetry and connectivity provide strong constraints between parts in a single shape; it also operates ``within category'', where global alignment helps part placement for shapes belong to same category. In contrast, \SCORES solves the shape composition problem; it works ``inter-shape'', ``cross-category'', and must handle much greater structural diversity and part discrepancies, making the problem technically challenging and the network difficult to structure and train.}

Computationally, \SCORES is built on iterative, hierarchical, substructure rectification based on a plausibility loss, while GRASS is a single pass without hierarchy resampling. Iterations are not needed for GRASS since it benefits from consistent global shape alignment. Further, \SCORES models the space of valid substructures with discrete latent space learning (VQ-VAE), while GRASS models the space of complete shapes. Our experiments show that \SCORES is significantly better than GRASS (and variants) for composition tasks. \rz{In a loose sense, the complexity gap between GRASS and \SCORES is analogous to global vs.~partial matching. It is well known that partial matching is considerably harder than global alignment.}

\vspace{-1mm}
\mypara{Generative shape priors.} There is a variety of methods that learn generative statistical models of shape spaces -- the survey of Mitra et al.~\shortcite{mitra2013} provides a good overview. In recent years, deep neural networks have been used to synthesize several shape representations, e.g. volumetric grids~\cite{girdhar2016,wu2015,wu2016}, point clouds~\cite{huang2015}, and part assemblies~\cite{li2017}. \SCORES is not, strictly speaking, a fully generative model: it is an iterative, structure-enhancing autoencoder which performs structure synthesis while conditioned on the input parts. Nevertheless, it shares a common context with the aforementioned priors, and is especially influenced by the last in terms of its recursive design. {\em Unlike} the other priors, however, it defines shape plausibility in terms of learned nested substructures, and can accommodate shapes that are globally unlike any seen during training but have plausible local structure.


\section{Method}
\label{sec:method}

\subsection{Problem statement and method design}
\SCORES takes as input a set of parts, typically from two or more different shapes, which
can be grossly misaligned, and have missing or redundant components. \SCORES converts this noisy soup of parts into a coherent, plausible, synthesized shape with all such defects corrected. It does not rely on part labels or other annotation to do so, and the final synthesized shape can look quite different from the input part collection -- adding, removing, and adjusting parts -- in order to maximize plausibility. By using an underlying shape prior to control the output, \SCORES goes beyond earlier methods that seek to simply fuse a fixed set of (typically heavily annotated) parts together.

\begin{figure}[t!] \centering
	\begin{overpic}[width=1.0\linewidth,tics=10]{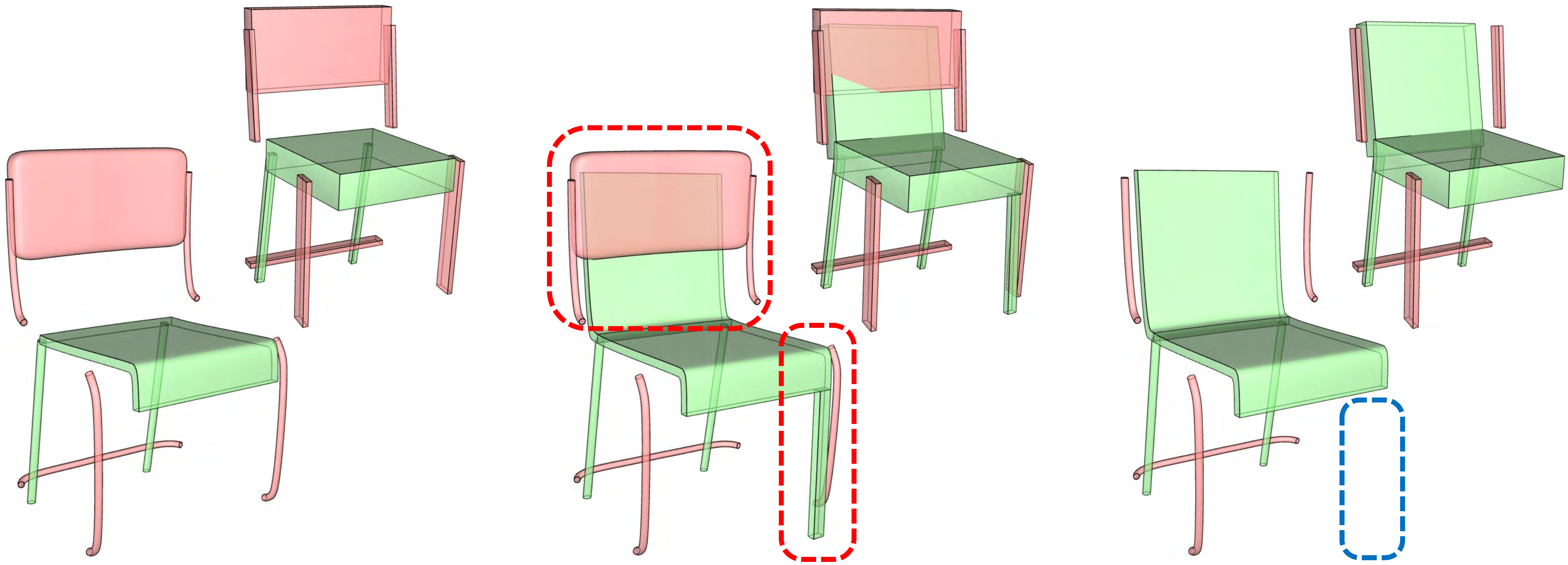}
    \put(16,-3){\small (a)}
    \put(50,-3){\small (b)}
    \put(85,-3){\small (c)}
	\end{overpic}
  \caption{Different types of noise in the input including alignment errors (a), redundant parts (marked with red dashed circles in (b)), and missing parts (indicated with a blue dashed circle in (c)).}
  \label{fig:noise_types}
  \vspace{-5mm}
\end{figure}

\vspace{-2mm}
\mypara{Input.}
The input to our method is a set of parts abstracted as simple oriented bounding boxes (OBBs). These parts are assumed to come from two or more different shapes, roughly aligned to a common coordinate system. \sid{In our experiments, we worked with consistently oriented and scaled shapes whose centroids were aligned for this initial placement.} This mixture commonly has three types of noise: alignment errors, redundant parts, and missing parts, as illustrated in Figure \ref{fig:noise_types}.

\vspace{-1mm}
\mypara{Output.}
The output of our method is a set of parts that form a coherent shape while maintaining the overall features and layout of the input parts as much as possible.
The output is generated in two stages: the first rectifies the noisy and misaligned input boxes into a plausible, connected layout; the second transforms the underlying parts based on their corresponding box adjustments. When a new box is generated, its underlying part geometry is retrieved from the shape database, following the contextual retrieval of Li et al.~\shortcite{li2017}.

\vspace{-1mm}
\mypara{Method design philosophy.}
Since the input parts may come from different shapes, merging them into a coherent structure faces key challenges. First, it is very likely there is no observed shape structure in the training set that completely explains the input set of parts. We must thus simultaneously infer a structural representation to accommodate the topology of the part relations, and improve the connections between parts by adjusting their geometry. Structure inference and geometry adjustment are coupled problems, which are not best addressed with a one-pass solution. Second, due to incompatibility among input parts, single-level global adjustment over the entire structure is likely to fail.

To address these challenges, we follow three principles:
\begin{itemize}
  \item \emph{Local adjustment.} Although the full set of input parts may not resemble any training structure, it is likely that its subsets form substructures which {\em are} echoed in exemplars~\cite{zheng2014}. Hence, these {\em substructures}, at various scales, should be adjusted with reference to valid substructures.
  \item \emph{Hierarchy-guided adjustment.} Randomly sampling substructures to adjust is neither efficient nor effective. Instead, we parse the input parts into a hierarchy, and use this hierarchy to sequentially adjust nested substructures in a top-down, cascading fashion.
  \item \emph{Iterative solution.} The structural organization and individual geometry of the parts should be inferred in an interleaved manner through iterative optimization.
\end{itemize}

\sid{
\mypara{Method overview.}
Based on these considerations, we formulate an iterative optimization which interleaves {\em hierarchy inference} and hierarchy-guided {\em local adjustment}. Given two sets of parts, \SCORES composes them through iterating over two key stages.
First, the parts are organized into a common hierarchy which best reflects their current layout. Second, the position and orientation of each part is adjusted to increase the plausibility of the assembly's constituent substructures, i.e. the subtrees of the inferred hierarchy. This stage itself comprises two passes. First, a bottom-up structure encoding pass assigns representative codes to each node of the hierarchy, capturing the arrangement of parts in its subtree, i.e. its substructure, in a context-free manner. Second, a top-down structure decoding pass adjusts these codes to correspond to more plausible part arrangements. The code adjustment is conducted with the help of a pre-learned manifold of valid substructures. Both the encoding and decoding functions are recursive neural networks trained by iteratively composing parts from exemplar shapes. Figure~\ref{fig:overview} provides an overview of the pipeline.
}


\IncMargin{0.5em}
\begin{algorithm}[b!]\small
\caption{Training and testing of SCORES.}
\label{algo:train}
\SetCommentSty{textsf}
\SetKwInOut{AlgoInput}{Input}
\SetKwInOut{AlgoOutput}{Output}
\SetKwFunction{TrainVQVAE}{TrainVQVAE}
\SetKwFunction{TrainDAE}{TrainDAE}
\SetKwFunction{AddNoise}{AddRandomNoise}
\Indm
\tcp{(OBB encoder/decoder $f^\text{box}_\text{enc}, f^\text{box}_\text{dec}$ omitted for clarity.)\vspace{1mm}}
\tcp{Training of SCORES: VQ-VAE and DAE}\label{cmt}
\Indp
\AlgoInput{ Training shape set: $\mathcal{S}=\{\mathcal{B}_i,\mathcal{H}_i\}_{i=1}^N$. }
\AlgoOutput{ DAE encoder: $f^\text{in}_\text{enc}$; DAE decoder: $f^\text{out}_\text{dec}$~(deform) and $f^\text{gen}_\text{dec}$~(synthesis); VQ-VAE codebook: $C$. }
$f^\text{in}_\text{enc}$, $f^\text{out}_\text{dec}$, $C$ $\leftarrow$ \TrainVQVAE{$\mathcal{S}$}\;\label{al:trainvqvae}
$\tilde{\mathcal{S}}$ $\leftarrow$ \AddNoise{$\mathcal{S}$}\;
$f^\text{in}_\text{enc}$, $f^\text{out}_\text{dec}$, $f^\text{gen}_\text{dec}$ $\leftarrow$ \TrainDAE{$\tilde{\mathcal{S}}$, $\mathcal{S}$, $C$}\;\label{al:traindae}
\Return $f^\text{in}_\text{enc}$, $f^\text{out}_\text{dec}$, $f^\text{gen}_\text{dec}$ and $C$\;
\SetCommentSty{textsf}
\SetKwInOut{AlgoInput}{Input}
\SetKwInOut{AlgoOutput}{Output}
\SetKwFunction{InitHier}{InitializeHierarchy}
\SetKwFunction{SampleHier}{HierarchyInference}
\SetKwFunction{Adjust}{LocalAdjustment}
\SetKwFunction{PartGeom}{PartGeometry}
\BlankLine
\Indm
\tcp{Optimization by testing SCORES}\label{cmt}
\Indp
\AlgoInput{ Trained networks/codebook: $f^\text{in}_\text{enc}$, $f^\text{out}_\text{dec}$, $f^\text{gen}_\text{dec}$ and $C$; \\ $K$ groups of part OBBs: $\mathcal{B} = \cup_{k=1}^K \mathcal{B}_k$. }
\AlgoOutput{ Fused shape : $S$. }
$\mathcal{H}$ $\leftarrow$ \InitHier{$\mathcal{B}$}\;
\Repeat {Stop condition is met} {
    $\mathcal{H}$ $\leftarrow$ \SampleHier{$C$, $\mathcal{B}$, $\mathcal{H}$}\;\label{al:hier}
    $\mathcal{B}$ $\leftarrow$ \Adjust{$f^\emph{in}_\emph{enc}$, $f^\emph{out}_\emph{dec}$, $f^\emph{gen}_\emph{dec}$, $C$, $\mathcal{B}$, $\mathcal{H}$}\;\label{al:adjust}
}
$S$ $\leftarrow$ \PartGeom{$\mathcal{B}$, $\mathcal{H}$}\tcp*[r]{Transform or generate parts}
\Return $S$\;
\end{algorithm}
\DecMargin{0.5em}

\subsection{Iterative optimization}

\mypara{Objective.}
To merge a set $\cB$ of part OBBs, we organize them into an optimal hierarchy $\cH$ whose subtrees are well explained by a learned substructure model. This objective is formulated as follows:
\vspace{-2.5mm}
\begin{equation}\label{eq:objective}
  \argmin_{\cH, \cB}{\sum_{n \in \cH}{\cE_\text{adjust}(\cB(S_n))}},
\vspace{-0.5mm}
\end{equation}
where $S_n$ is the substructure rooted at node $n$, and $\cB(S_n) \subset \cB$ is its set of OBBs. $\cE_\text{adjust}$ is an adjustment energy which measures the deviation of a substructure from valid counterparts. Such deviation is minimized by local adjustment of the substructure.

\vspace{-0.5mm}
\mypara{Interleaving optimization.}
The above objective is optimized iteratively. Each iteration comprises two distinct steps: (a) {\em hierarchy inference} and (b) {\em substructure adjustment}. The first step computes a plausible hierarchical organization (parse tree) for the current set of part boxes. The second step adjusts the geometry of the boxes based on \SCORES's {\em recursive denoising autoencoder}, trained to optimize the box configuration according to a {\em learned structure prior}. While the encoding phase of \SCORES aggregates structural information hierarchically, the decoding phase performs a cascading adjustment of the substructures in the hierarchy, based on the learned model of valid substructures (see Figure~\ref{fig:overview}(a,b)). The latter may sometimes regenerate a substructure, thereby adding/removing boxes, if necessary. The process is guaranteed to converge if both steps decrease the energy (discussion in supplementary material).

The step-by-step training and optimization (testing) processes are shown in Algorithm~\ref{algo:train}.
\cy{We first learn the model (prior) of valid substructures (Line~\ref{al:trainvqvae}; Section~\ref{subsec:vqvae}) and then
train the encoder-decoder networks for substructure adjustment (Line~\ref{al:traindae}; Section~\ref{subsec:dae}).
The learned substructure prior and the trained networks are used for iterative optimization (Line~\ref{al:hier} and~\ref{al:adjust}; Section~\ref{subsec:hier} and~\ref{subsec:adjust}).} Below, we describe these phases in detail.

\subsection{Training: Substructure prior and denoising network}
\label{subsec:prior}

We train our structure fusion model in two stages. First, we learn a codebook for valid substructures at various levels of abstraction. This stage is trained with clean, ground-truth shapes. Second, we fine-tune the neural network associated with the (frozen) codebook, in order to denoise part assemblies corrupted by synthetic noise.

\subsubsection{Discrete latent space of substructures}
\label{subsec:vqvae}

\begin{figure}[t!] \centering
	\begin{overpic}[width=1.0\linewidth,tics=10]{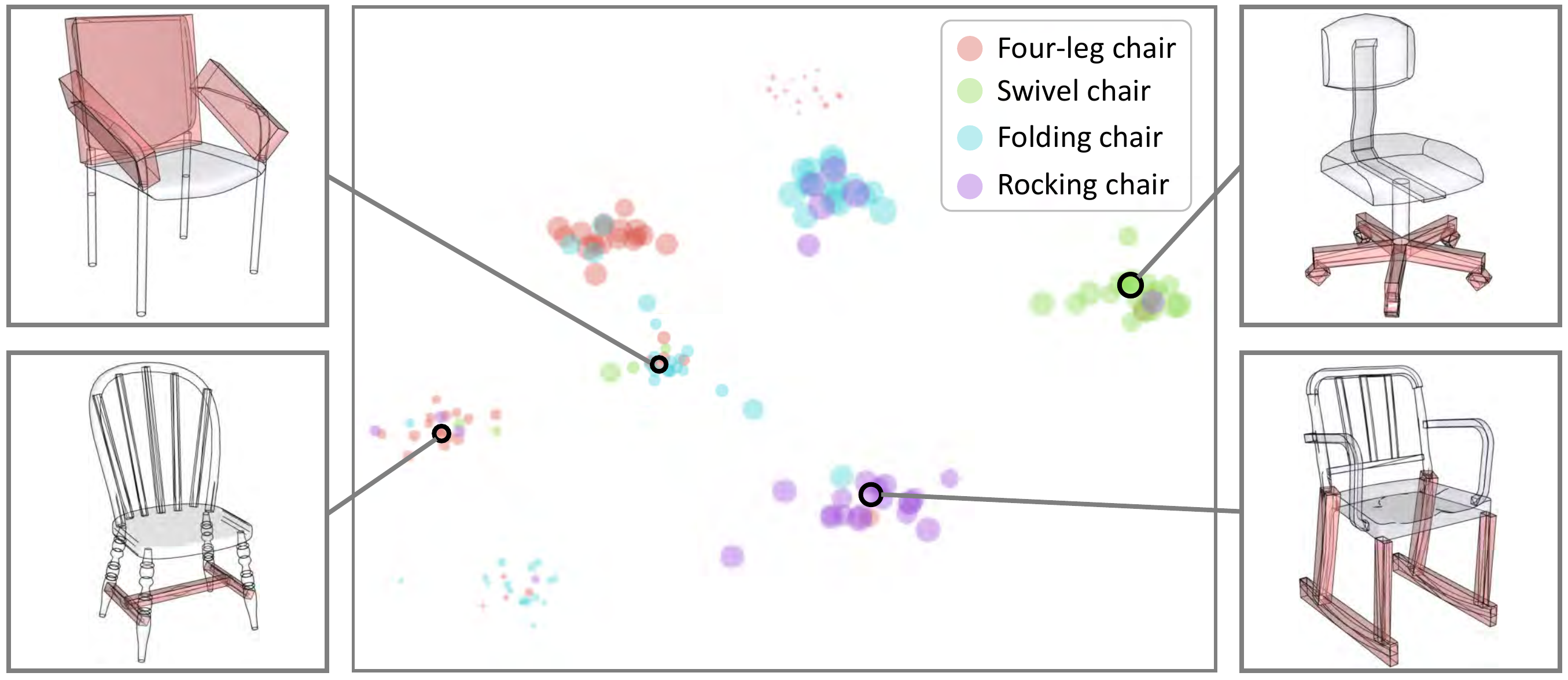}
	\end{overpic}
    \caption{A t-SNE visualization of outer code embedding (latent space) of substructures (indicated by dots). The substructures are sampled from the hierarchies of chairs in four categories (color-coded), including four-leg, swivel, folding and rocking chairs. The size of a dot indicates the number of parts of a substructure. Four representative substructures, corresponding to codebook vectors in the discrete latent space, are shown on the sides.}
    \label{fig:embedding}
    \vspace{-7pt}
\end{figure}

Each topologically and functionally valid substructure is a discrete and isolated point in the space of all part layouts, since every part in it needs to fit just so and random perturbations can break the careful arrangement. More accurately, each substructure represents a small family of variants, generated by highly correlated changes to its parameters that preserve connectivity and other functional properties, that lie on a low-dimensional local manifold. Each such family of local variants can be considered a mode of the highly multi-modal distribution of plausible substructures.
Therefore, we model the space of valid substructures via learning a feature embedding of substructures sampled from exemplar shapes (see Figure~\ref{fig:embedding}).

Instead of modeling randomly sampled substructures, we encode part structures of complete exemplar shapes with recursive neural networks (RvNNs)~\cite{li2017}. The resulting part hierarchy contains a cascade of nested valid substructures as subtrees. A major benefit of analyzing substructures in complete shape hierarchies is that it allows us to learn direct contextual dependencies between different substructures.

\vspace{-1mm}
\mypara{Context-enhanced substructure encoding.}
To learn a space of valid substructures, we must first embed all substructures in the part hierarchies of training shapes to a common feature space. To this end, \cy{we propose a recursive autoencoder~\cite{socher2011} which recursively generates fixed-dimensional codes for each node of the hierarchy representing a substructure.} This is achieved with a bottom-up encoding for structural information aggregation, followed by a top-down decoding for contextual information propagation~\cite{le2014}.

\begin{figure}[t!] \centering
	\begin{overpic}[width=1.0\linewidth,tics=10]{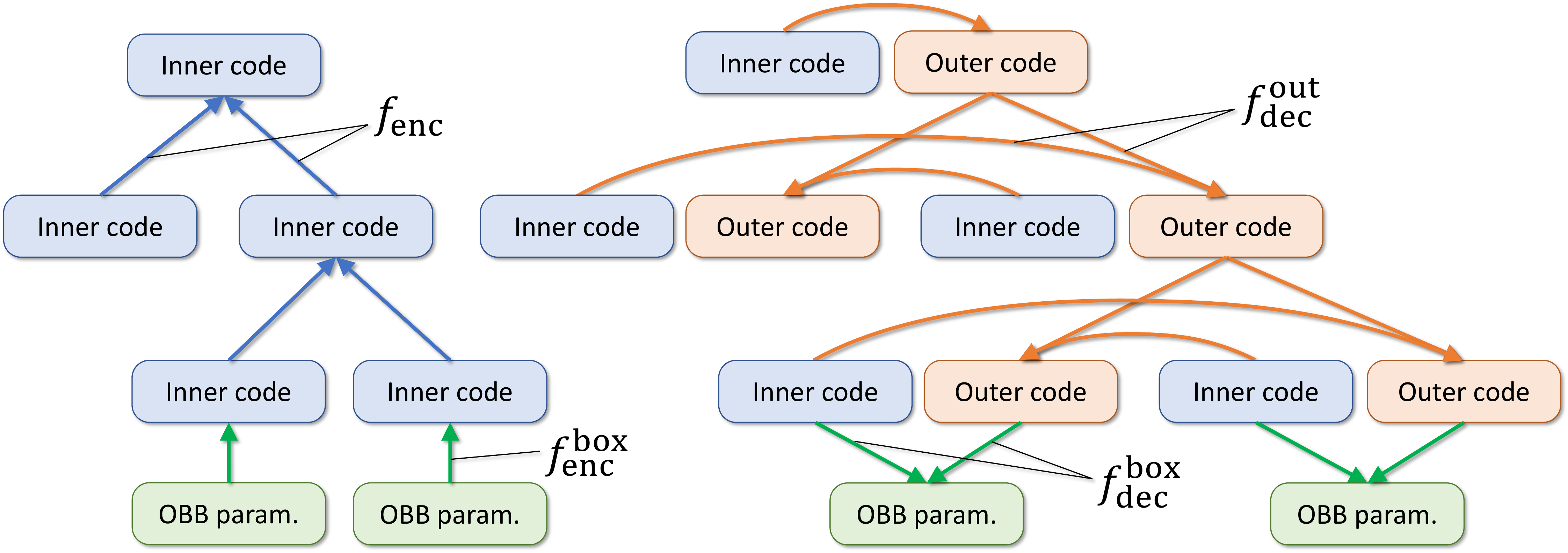}
    \put(2,-3){\small (a) Inner code generation.}
    \put(49,-3){\small (b) Outer code generation.}
	\end{overpic}
    \caption{An illustration of bottom-up encoding of inner codes and top-down decoding of outer codes.}
    \label{fig:inout}
    \vspace{-7pt}
\end{figure}

The bottom-up encoding generates an \emph{inner code} for each node, which captures the {\em local geometry} of the subtree rooted there. Inner codes of child nodes are concatenated and fed to a shared fully-connected module to yield their parent's inner code (Figure \ref{fig:inout}(a)):
\begin{equation}\label{eq:inner}
  x_p^\text{in} = f^\text{in}_\text{enc}(x_l^\text{in}, x_r^\text{in}),
\end{equation}
where $x_l^\text{in}$, $x_r^\text{in}$ and $x_p^\text{in}$ denote the inner codes of two sibling nodes and their parent node, respectively. $f^\text{in}_\text{enc}$ is a multi-layer perceptron (MLP) with two hidden layers, which encodes either adjacency or symmetry-based grouping~\cite{li2017}.

The top-down decoding produces an \emph{outer code} for a node, through decoding from its parent's outer code, as well as the inner codes of its siblings (Figure \ref{fig:inout}(b)):
\begin{equation}\label{eq:outer}
  x_i^\text{out} = f_\text{dec}^\text{out}(x_p^\text{out}, x_s^\text{in}),
\end{equation}
where $x_i^\text{out}$, $x_p^\text{out}$ and $x_s^\text{in}$ denote the outer codes for a node and its parent, and the inner code of a sibling, respectively. $f_\text{dec}^\text{out}$ is a two-layer MLP decoder, again corresponding to either adjacency or symmetry-based ungrouping.
\kx{A node classifier is trained to determine whether a node represents adjacency or symmetry grouping~\cite{li2017}.}
The principal role of the outer code is to capture a node's {\em global context} in the overall layout.

An additional {\em box encoder} generates the initial codes for part OBBs (input to the bottom-up pass) from box features, and a {\em box decoder} generates the final adjusted OBBs from the concatenated inner and outer codes of the leaf nodes after the top-down pass:
\begin{equation}\label{eq:box}
  x_b^\text{in} = f_\text{enc}^\text{box}(b_i), \quad b_o = f_\text{dec}^\text{box}(x_b^\text{in}, x_b^\text{out}),
\end{equation}
where $x_b^\text{in}$ and $x_b^\text{out}$ denote the inner and outer codes for a leaf node, respectively.
$b_i$ and $b_o$ are the parameter vectors of input and output OBB, respectively.
$f_\text{enc}^\text{box}$ and $f_\text{dec}^\text{box}$ are two-layer MLPs for box encoding and decoding, respectively.
Please refer to supplementary material for details on hyper-parameters.

We made some design decisions in our treatment of inner and outer codes. For instance, we could have chosen to condition outer code generation on the concatenated inner and outer codes of the parent, or on the inner code of the node itself (and not just its sibling). We experimented with these alternatives. Concatenation yielded similar results, but slower convergence. Conditioning on the node's own inner code also did not affect results, but we chose to avoid it to force the outer code to learn meaningful structure on its own. We did use both inner and outer codes for final OBB reconstruction, in order to achieve a conservative adjustment that respects the input.


\mypara{A codebook of valid substructures.}
For each internal node representing a valid substructure, we use its \emph{outer} code as a globally contextualized feature representation and learn a latent space of such feature based on a variational autoencoder (VAE)~\cite{doersch2016vae}. Since the subspace for valid substructures is discrete in nature, we model it using discrete latent representations, specifically a vector quantized-variational autoencoder (VQ-VAE) \cite{oord2017}. This approach combines VAE training with a dictionary learning: the latent code for an input is ``snapped'' to the nearest entry in a jointly learned finite codebook before being passed to the decoder. An input datapoint $x$ is represented in a VQ-VAE with learned codebook vectors $C = \{e_i\}_{i=1}^K$ as
\vspace{-0.5mm}
\[
z_q(x) = e_k, \quad \text{where} \quad k = \argmin_i\|z_e(x) - e_i\|_2.
\vspace{-2mm}
\]
where $z_e(x)$ is the encoder output.

\begin{figure}[t!] \centering
	\begin{overpic}[width=1.0\linewidth,tics=10]{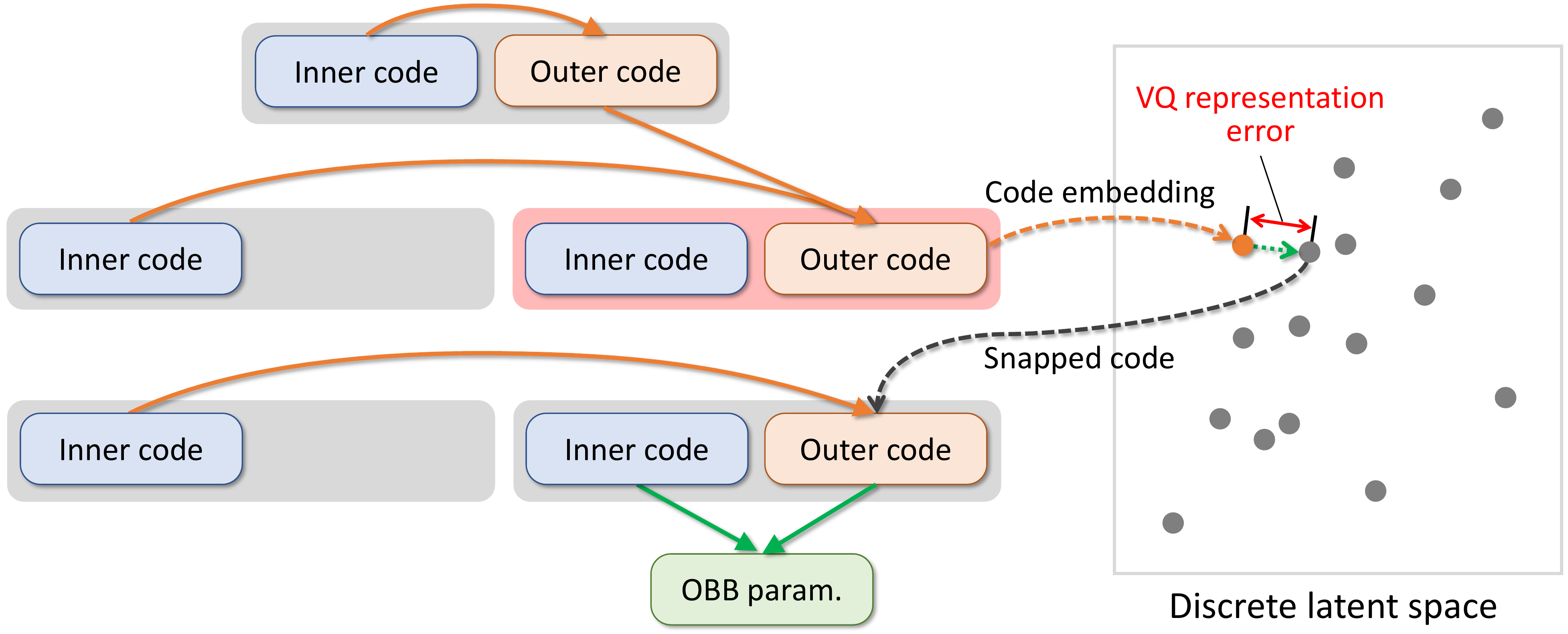}
	\end{overpic}
    \caption{Local adjustment of the substructure corresponding to the internal node shaded in red. The adjustment is conducted in a top-down pass, with outer code computation (embedding) and nearest neighbor snapping (demonstrated only for the right child) in a discrete latent space (rightmost) learned from valid substructures. The VQ representation error, which is the distance between embedded and snapped codes, measures how much the substructure deviates from a valid one.}
    \label{fig:objective}
    \vspace{-7pt}
\end{figure}

The VQ-VAE is trained on substructures sampled from hierarchies of training shapes. (These hierarchies are computed by GRASS~\cite{li2017}.)
The details on learning the encoder/decoder, as well as the VQ codebook, can be found in supplementary material.
Figure~\ref{fig:embedding} visualizes the learned discrete latent space of valid substructures. Once trained, we can define for a given substructure $S_n$ rooted at node $n$ a VQ representation error, i.e. its deviation from the learned model of valid substructures:
\vspace{-0.5mm}
\begin{equation}\label{eq:reperror}
  \cE_\text{VQ}(S_n) = \| z_q(x^\text{out}_n) - x^\text{out}_n \|_2^2,
\vspace{-0.5mm}
\end{equation}
where $x^\text{out}_n$ is the outer code of node $n$. Figure~\ref{fig:objective} illustrates the VQ representation of a substructure via nearest neighbor search in a discrete latent space.


Consequently, the objective in Equation (\ref{eq:objective}) can be written as the overall VQ error at all internal nodes in hierarchy $\cH$:
\vspace{-0.5mm}
\begin{equation}\label{eq:objvq}
  \argmin_{\cH, \cB}{\sum_{n \in \cH}{\| z_q(x^\text{out}_n) - x^\text{out}_n \|_2^2}}.
  \vspace{-0.5mm}
\end{equation}

\subsubsection{Fine-tuning the autoencoder for denoising}
\label{subsec:dae}

The VQ-VAE codebook $C$ is trained on clean, ground-truth shapes, in order to learn the correct substructure prior. Now, holding this codebook frozen, we will fine-tune the encoder-decoder networks to repair {\em noisy} part assemblies, which is the main goal of this paper. We select random subtrees of training shape hierarchies as substructures to be denoised. We then add synthetic noise to them, to mimic the effect of creative assembly (details in supplementary material): the original substructure is the desired denoised output.

To perform the denoising, we re-use the recursive encoder-decoder networks used for substructure feature learning to create a {\em denoising autoencoder} (DAE), which performs cascading substructure adjustment guided by the current part hierarchy. Taking the current set of part OBBs as input, the DAE performs an RvNN encoding followed by an RvNN decoding, where the inner and outer codes for each node are computed as above. The key feature of the DAE is substructure adjustment performed at each internal node during the decoding phase, based on its outer code. The adjustment relies on the learned latent space of valid substructures. Specifically, we reduce the VQ representation error for an internal node by VQ-VAE denoising~\cite{oord2017} -- ``snapping'' its outer code to the nearest codebook vector -- and continue decoding with the snapped outer code (see Figure~\ref{fig:objective} for an illustration of this process). After a full pass of top-down decoding is finished, a denoised configuration of OBBs is generated.

This training step results in updated networks $f^\text{in}_\text{enc}, f^\text{box}_\text{enc}, f^\text{out}_\text{dec}$ and $f^\text{box}_\text{dec}$ which can map even noisy substructures to outer codes close to the corresponding ``clean'' codebook entries. VQ-VAE ``snapping'' then handles the necessary residual denoising. These fine-tuned DAE networks, plus the VQ-VAE codebook, constitute the SCORES model for structure fusion via local adjustment.

In cases where a part of the input is too noisy to be fixed by local adjustment alone (e.g. parts must be added or removed), we fully re-synthesize this substructure using a jointly trained {\em synthesizing decoder} $f^\text{gen}_\text{dec}$, \cy{details can be found in supplementary material}. For ease of presentation, this component is discussed separately below.

\subsection{Testing: Iteratively optimizing a noisy part assembly}

Our structure fusion procedure proceeds in two alternating steps, repeated until convergence.

\vspace{-1mm}
\subsubsection{Alternating step 1: Hierarchy inference}
\label{subsec:hier}
The first step takes the current set of part OBBs as input, and infers an encoding hierarchy for it so that the substructures rooted at internal nodes are best explained by the VQ-VAE substructure prior. Our task is to search for a hierarchy minimizing Equation (\ref{eq:objvq}), which is an NP-hard problem. To avoid exhaustive search, we resort to an importance sampling strategy. Proceeding top-down from the root node of the current hierarchy, we select a set of internal nodes whose subtree hierarchies will be resampled. The selection is based on VQ error: a node with high error is likely to benefit from resampling. Specifically, we select a node $n$, whose subtree is substructure $S_n$, for resampling with probability \mbox{$p_n = e^{-{\cE_\text{VQ}^2(S_n)}/\left(\sigma^2\cE_\text{max}^2\right)}$}, where $\cE_\text{max}$ is the maximum error among all internal nodes and $\sigma = 0.6$ by default. $p_n$ is set to 0 for the root node.

\begin{figure}[t!] \centering
	\begin{overpic}[width=1.0\linewidth,tics=10]{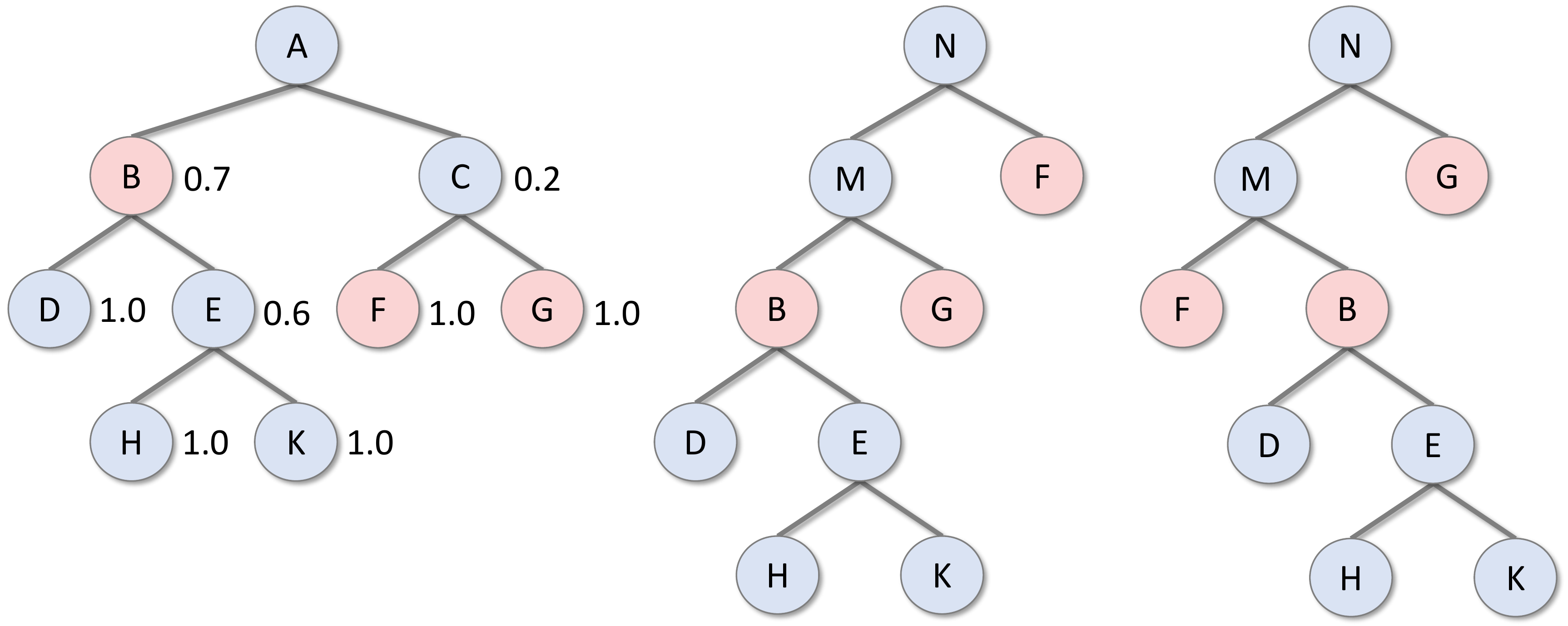}
    \put(18,5){\small (a)}
    \put(59,-3){\small (b)}
    \put(85,-3){\small (c)}
	\end{overpic}
    \caption{An illustration of hierarchy resampling. Given the hierarchy from the previous iteration (a), a sampling probability is estimated for each internal node (numbers beside nodes), based on its representation error (small error implies large probability). Once an internal node is selected (e.g. node `B'), its descendants will not be selected. The selected nodes (red) are used to resample new hierarchies, with random permutations (only two are shown in (b) and (c)). In the new hierarchies, the subtree under `B' from the old hierarchy is transplanted verbatim.}
    \label{fig:hierinfer}
    \vspace{-7pt}
\end{figure}

For each selected node, we randomly resample at most $M = 10$ new hierarchical groupings of the leaf nodes in its subtree. The whole process is repeated $N = 10$ times. Figure~\ref{fig:hierinfer} shows an illustration of hierarchy resampling. If a hierarchy with smaller overall VQ error is found, it is used for the next iteration, else the previous hierarchy is retained.

To bootstrap the iteration, an initial hierarchy is constructed by inducing a tree for the parts from an individual source shape, as a subgraph of the shape's reference hierarchy computed using GRASS~\cite{li2017}. These subtrees are then linked together with a common parent. When the source shape or its hierarchy is unknown, we simply build a hierarchy with all input OBBs from scratch using, again, the method in GRASS.

In Figure~\ref{fig:iteradjust}(bottom), we show the effect of not recomputing the hierarchy after each local adjustment step. Without recomputation, correctly grouped substructures are not available for adjustment: the method converges slower and to worse output.

\subsubsection{Alternating step 2: Local adjustment of substructures}
\label{subsec:adjust}
The second alternating step of structure fusion is {\em local adjustment}, in which part OBB shapes and positions are adjusted for greater plausibility and coherence, guided by the inferred hierarchy. A straightforward approach would be a denoising version of GRASS~\cite{li2017}: the part assembly is recursively encoded to a root code, and decoded back to a ``clean'' version, with training comprising noisy/clean shapes. But as noted above, this method fails for structure fusion. GRASS-type methods work when the desired output is globally similar to training shapes. However, for fusion, the input assemblies could be quite different from anything seen in training, with similarity only at the local level.

Hence, we use a prior over plausible substructures, rather than complete shapes alone. The recursive denoising autoencoder (DAE) learned in Section \ref{subsec:prior}, guided by the VQ-VAE substructure prior, is used to adjust the OBBs towards greater plausibility over local contexts. The part OBBs of the noisy assembly are mapped to inner codes by the box encoder $f^\text{box}_\text{enc}$; the inner codes are propagated up the hierarchy inferred above by recursively applying the encoder network $f^\text{in}_\text{enc}$; outer codes are generated top-down using the decoder network $f^\text{out}_\text{dec}$ and the the VQ-VAE codebook $C$; and finally the denoised codes at the leaf level are mapped back to clean OBBs using the box decoder $f^\text{box}_\text{dec}$.

\begin{figure}[t!] \centering
	\begin{overpic}[width=1.0\linewidth,tics=10]{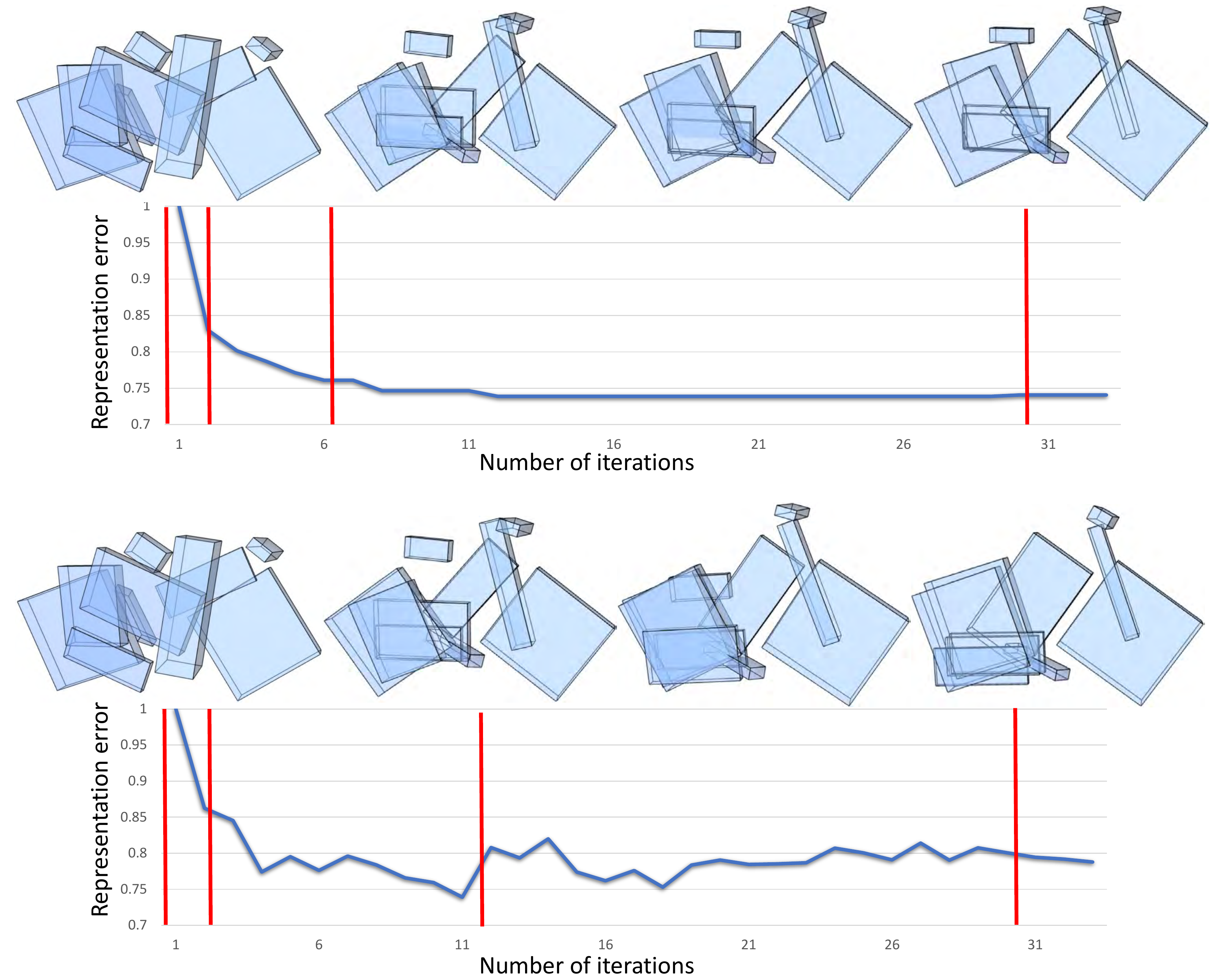}
	\end{overpic}
    \caption{Change of representation error with iterative part adjustment, with (top row) and without (bottom row) hierarchy resampling. Given a randomly perturbed part configuration of a bicycle model, we show the results of adjustment at various iteration steps (marked with red vertical lines). Our method converges better with hierarchy resampling.}
    \label{fig:iteradjust}
    \vspace{-7pt}
\end{figure}

Figure~\ref{fig:iteradjust}(top) shows the decrease in overall VQ error as the iterative optimization proceeds.

\mypara{Deformation vs. Synthesis.} Aggregating parts from multiple different shapes may lead to redundant or missing parts (Figure \ref{fig:noise_types}). This can yield a substructure with large VQ error, which code snapping is unable to correct since it cannot change part counts. In this case, a denoising autoencoder that entirely re-synthesizes the substructure from its root code can fix the defect, at the loss of some fidelity to the input geometry. However, as we noted earlier, such an approach produces shapes similar to training exemplars. If applied at global scale, it is unsuited for open-ended design tasks where arbitrary and novel shapes may be created.

\begin{figure}[t!] \centering
	\begin{overpic}[width=1.0\linewidth,tics=10]{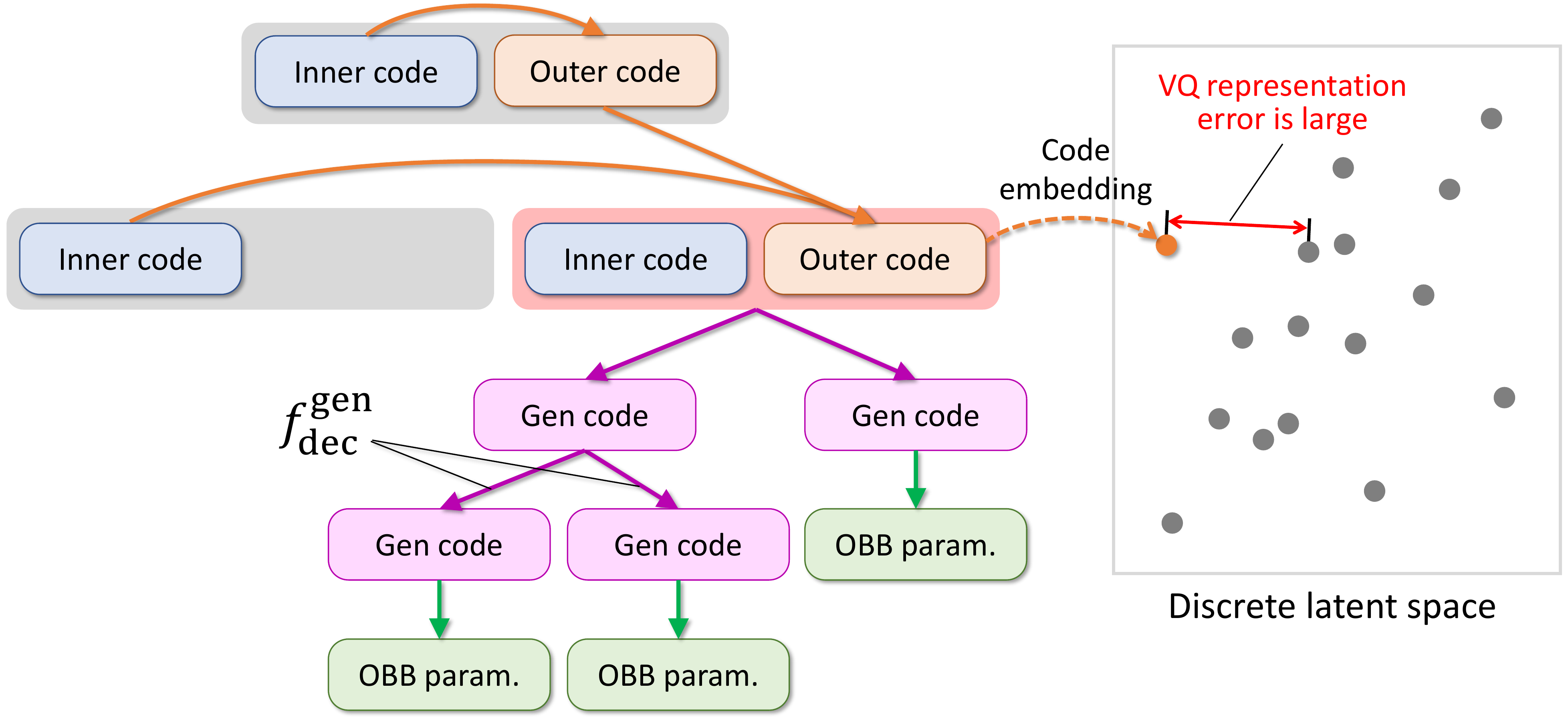}
	\end{overpic}
    \caption{Illustration of full re-synthesis of the substructure corresponding to the internal node shaded in red. Since the VQ representation loss is too large, the substructure is deemed implausible, and is recreated from scratch with the synthesis decoder $f_\text{dec}^\text{gen}$. This decoder is trained for repairing substructures with redundant or missing parts.}
    \label{fig:fullgen}
    \vspace{-3mm}
\end{figure}

Hence, our local adjustment includes an exploratory tradeoff between \emph{geometric deformation} (that preserves the number of input parts) and \emph{full synthesis} (which may add or remove parts). Our metric is naturally the VQ error at each node. If the metric is small, we accept it as a plausible substructure and merely seek to improve its geometry via VQ-VAE based denoising, without changing its part composition. Otherwise, we re-synthesize it from scratch 
from its concatenated inner and outer codes:
\begin{equation}\label{eq:gen}
  [x_l^\text{gen},x_r^\text{gen}] = f_\text{dec}^\text{gen}(x_p^\text{gen}),
\end{equation}
where $x_p^\text{gen}$, $x_l^\text{gen}$ and $x_r^\text{gen}$ are the codes of a parent node and its two children, respectively, and $f_\text{dec}^\text{gen}$ is a two-layer MLP. The root node of the substructure is assigned the code \mbox{$x_n^\text{gen} = [x_n^\text{in},x_n^\text{out}]$}. See Figure~\ref{fig:fullgen} for an illustration of full synthesis of a substructure.

The threshold determining when the error is large enough for full synthesis is an interactive, user-specified parameter. To avoid easily triggering re-synthesis of globally unusual part combinations (high error at the root node), we add node depth to the metric, suppressing re-synthesis of larger substructures:
\begin{equation}\label{eq:metric}
  \eta(S_n) = \left(\frac{\cE_\text{VQ}(S_n)}{\cE_\text{max}}\right)^\alpha \left(\frac{d_n}{d_\text{max}}\right)^{1-\alpha}
\end{equation}
where $d_n$ is the depth of node $n$ and $d_\text{max}$ the maximum depth of the hierarchy. We use $\alpha=0.3$ for all experiments, and per-category fine-tuning of $\alpha$ could get better performance.



\subsubsection{Synthesizing fine-grained shape geometry}
\label{subsec:geometry}
Having obtained the assembled structure of part OBBs, we transform each input part to its output (denoised) placement by aligning input and output OBBs. If a subtree is re-synthesized, we fill newly created OBBs with database parts retrieved by SARF codes~\cite{li2017}. Since our local adjustment is over OBBs, transformed parts may have small misalignments at connection points. We improve these connections with least-squares optimization of closest ``docking points'' on adjacent parts~\cite{kalogerakis2012}. \SCORES already leads to a plausible assembly of parts, so this post-adjustment is sparse and minor: only ${\sim}10\%$ shapes required it. The amount of post-adjustment is no more than $2.2\%$ of the OBB diagonal of the entire shape for translation, and $3.5\%$ for scaling. This is typically less than $5\%$ of the scale of our RvNN-based substructure adjustment.



\vspace{3pt}
\kx{
Please refer to the source code on our project page\footnote{\url{http://kevinkaixu.net/projects/scores.html}} to clarify details and reproduce our results.
}

\section{Results and Evaluation}
\label{sec:results}

In this section, we evaluate the performance of \SCORES for structure fusion and show results on multiple shape categories.

\mypara{Dataset and benchmark.}
We tested our algorithm on the publicly available ComplementMe dataset~\cite{sung2017}, which is a subset of the ShapeNet repository~\cite{chang2015_shapeNet}.
Each shape has part segmentations for the purpose of assembly-based shape modeling. Like extensive prior work~\cite{mitra2013}, we rely on access to a segmented dataset, typically obtained by some combination of automatic and manual annotation. While the method can handle some over-segmentation, we cannot work with completely unsegmented shapes. We did not use any part labels throughout the training and testing processes: our method is {\bf label-free}.

Dataset statistics are given in Table \ref{tab:dataset}. For each category, 20\% of the shapes were chosen for testing, and the remaining 80\% for training. The parts in each test shape were randomly partitioned into groups, which will be mixed and matched to test structure fusion. Each shape contributes at most $12$ different partitions. The number of parts per group ranges in $[1,N-1]$ ($N$ is the shape's part count). We call this the {\sc StructMerge} benchmark, and will make this publicly available, along with all our code.

\begin{table}[t!]
  \begin{tabular}{c|c|c}
    {\bf Category} & {\bf \# Shapes} & {\bf \# Substructures} \\
    \hline
    \hline
    Airplane & $215$ & $4328$ \\
    Bicycle & $145$ & $8763$ \\
    Candelabrum & $100$ & $1017$ \\
    Chair & $277$ & $19391$ \\
    Lamp & $188$ & $2376$ \\
    Table & $176$ & $7620$
  \end{tabular}
  \caption{Dataset statistics. For each category, we list the shape count and the number of substructures sampled from the shape hierarchies.}
  \label{tab:dataset}
  \vspace{-7mm}
\end{table}

\begin{figure}[t!]
  \includegraphics[width=\linewidth]{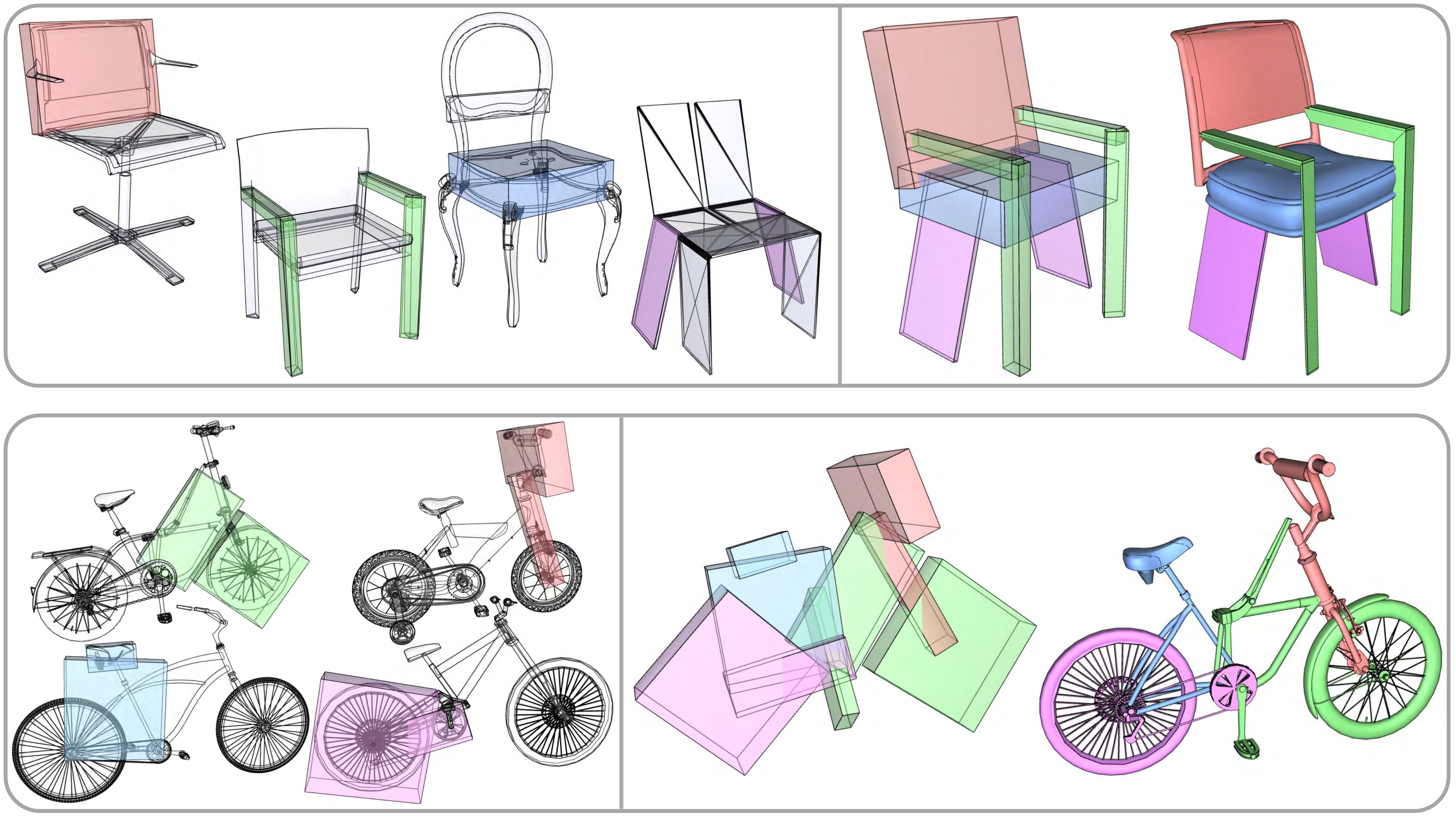}
	\vspace{-7mm}
  \caption{Merging parts from multiple source shapes.}
  \label{fig:multi_source}
	\vspace{-2mm}
\end{figure}

\begin{figure*}[t!]
  \includegraphics[width=\linewidth]{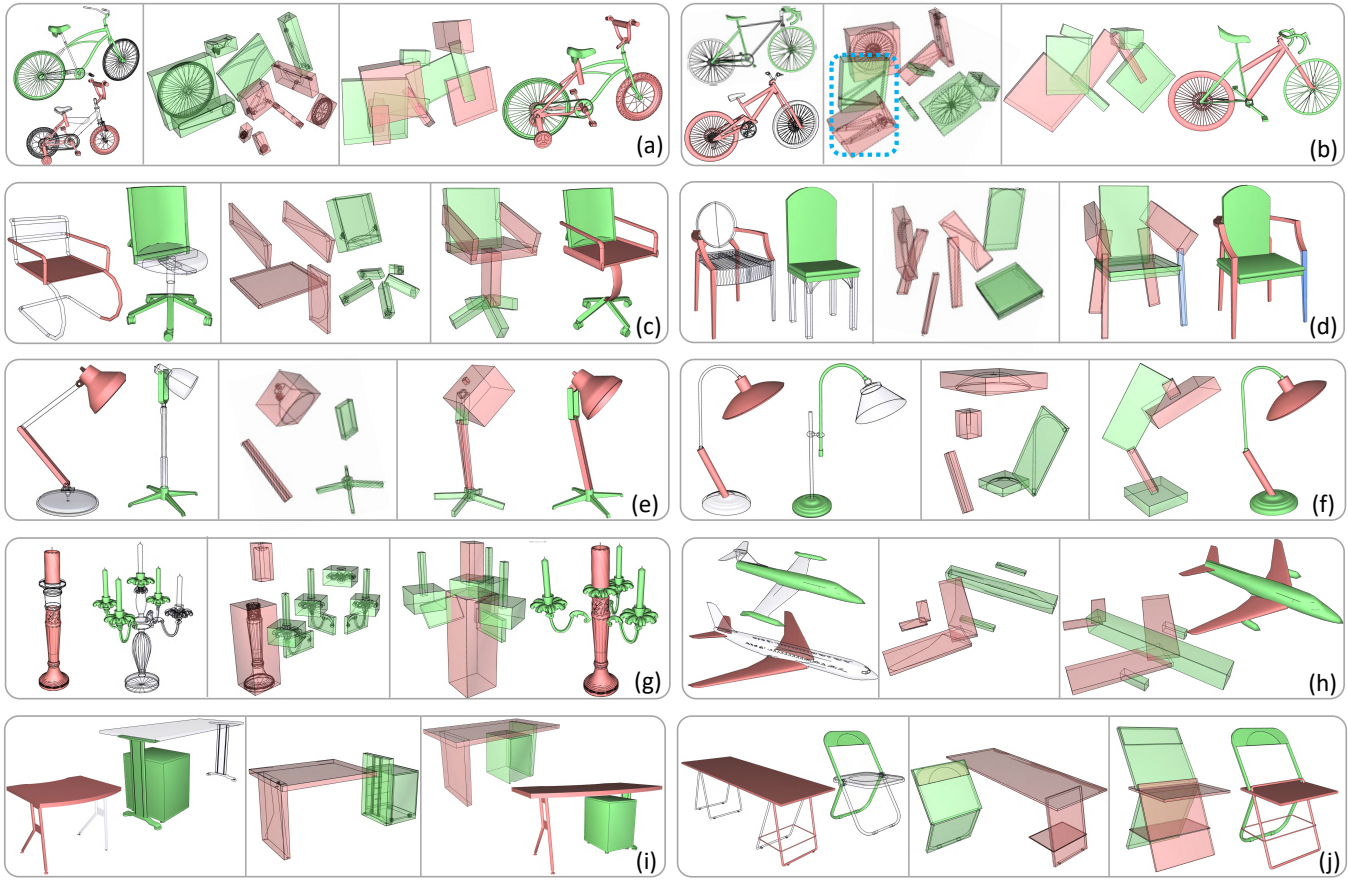}
  \caption{Visual examples of challenging compositions achieved by \SCORES over six object categories. In each block, the leftmost box shows parts selected from two source shapes for merging (red and green). The middle box shows the OBB inputs to the algorithm -- an initial hierarchy is computed for each group of parts separately (as a subgraph of the inferred source hierarchy) and they are then linked by a common parent. The rightmost box shows the final merged result (OBB and fine-grained geometry). Redundant parts in the sources are circled with blue dots (b). New parts synthesized to better fit the shape prior are colored in blue (d). In (j), we show cross-category part merging between Chair and Table. Substructure adjustment was conducted based on the substructure model (discrete latent space) learned for chairs, thus yielding a chair-like final shape.}
  \label{fig:hetero_visual}
  \vspace{-2mm}
\end{figure*}

We use this benchmark to test both local adjustment without structure synthesis, as well as structural composition that is free to judiciously add and remove parts to improve plausibility. In both cases, we employ an initial quantitative metric that measures whether two part groups from the {\em same} source shape, with added noise, can be successfully merged. (We discuss results for more than two part groups separately.) We test our method's robustness to varying noise levels in the evaluation below. The error in part merging is measured as the sum of squared L2 distances between the (noise-free) ground-truth part OBBs and the corresponding OBBs in the merged output. This metric of course does not reflect the algorithm's ability to merge parts from {\em different} source shapes, which is the significant application of our method. For this, we must resort to human visual evaluation, since no automatic test is suitable. In the tests below, we present both visual comparisons (Figures \ref{fig:teaser}, \ref{fig:multi_source} and \ref{fig:hetero_visual}) as well as the results from user experiments.

\begin{figure}[t!]
  \includegraphics[width=\linewidth]{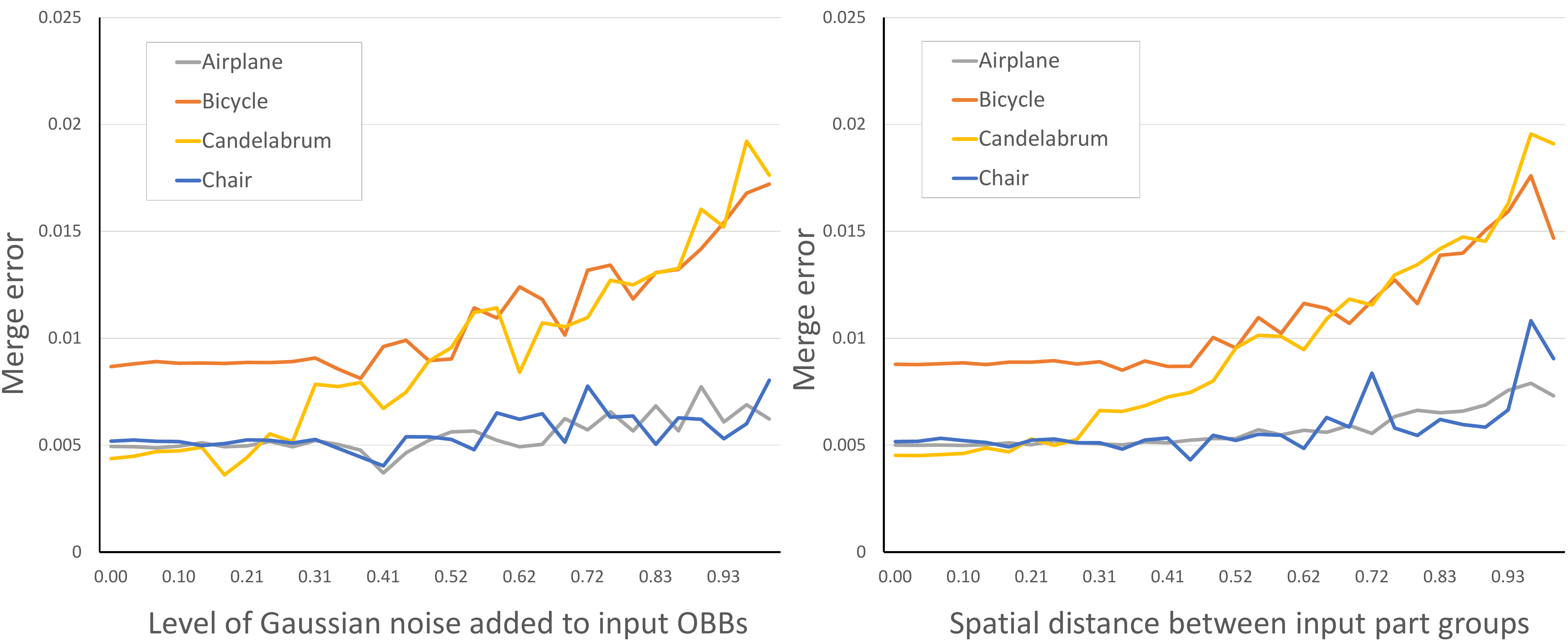}
  \vspace{-5.5mm}
  \caption{Performance (merge-error against ground-truth) of our method over varying amounts of perturbation, evaluated on shapes in four categories from our {\sc StructMerge} benchmark. (a): Increasing level of Gaussian noise is added to the input vector of OBB parameters. (b): Spatial distance between two part groups is increased. Noise level and spatial distance are measured w.r.t. the OBB diagonal of the entire shape.}
  \label{fig:exact-noise}
  \vspace{-3mm}
\end{figure}

\mypara{Exact composition (no synthesis).} We first present plots showing the performance of our method in merging part structures with only local adjustments of part placement and geometry. In Figure \ref{fig:exact-noise}, we show the post-merge error (vs ground truth) with noisy input OBBs, averaged over all shapes in our benchmark. To add noise, the vector encoding the parameters (positions, axes and dimensions) of an input OBB is perturbed with Gaussian noise. The plots in \ref{fig:exact-noise}(a) show the error for varying noise levels. For a more intuitive demonstration of the robustness of our method to initial misalignment of part groups, we plot in \ref{fig:exact-noise}(b) the merge error over varying spatial distance between two groups. The distance is measured with respect to the OBB diagonal length of the full ground-truth shape.

\begin{figure}[ht!]
  \includegraphics[width=\linewidth]{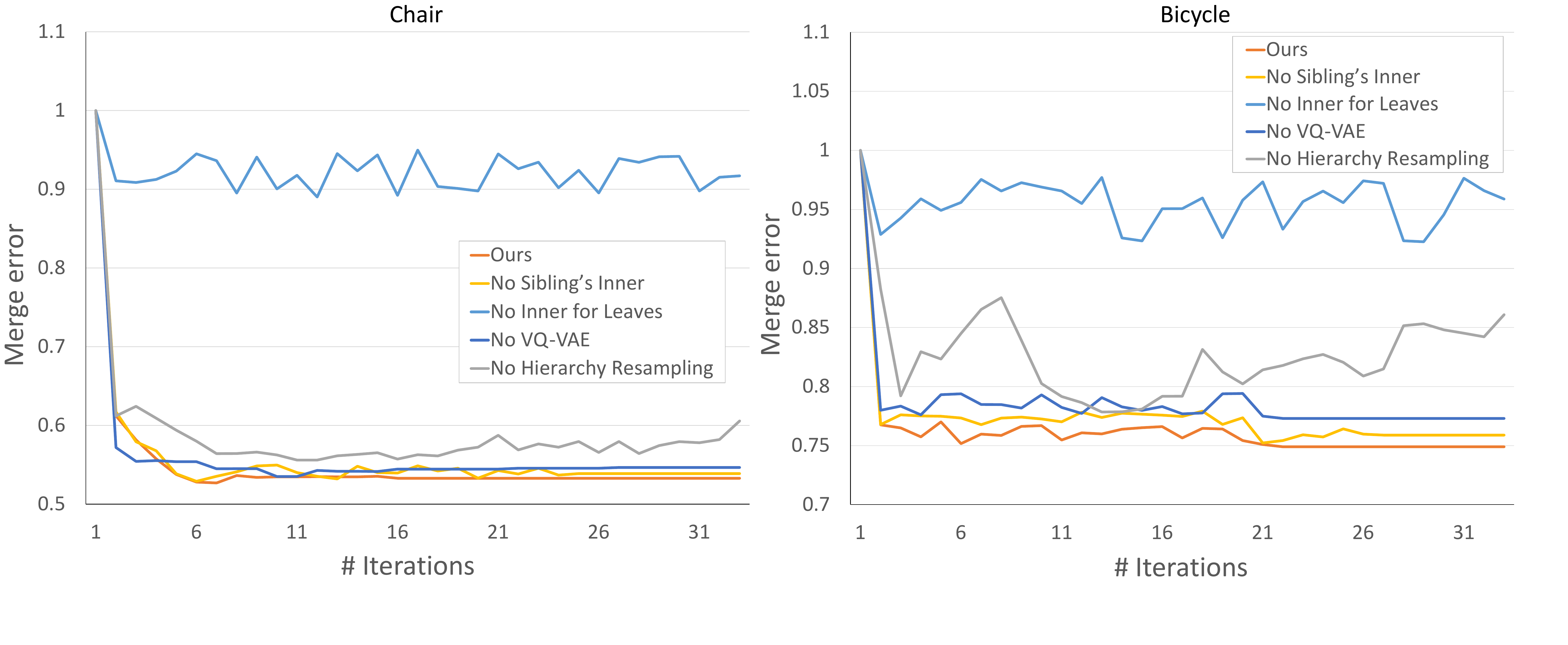}
  \vspace{-9mm}
  \caption{Performance (merge-error) of our method vs alternatives for fusing structures with local adjustments only, over an increasing number of iterations. The evaluation is conducted with shapes in two categories (Chair and Bicycle) from our {\sc StructMerge} benchmark.}
  \label{fig:exact-ablation}
  \vspace{-3mm}
\end{figure}

In Figure~\ref{fig:exact-ablation}, we show a quantitative comparison for several ablated alternative methods, namely:
\begin{itemize}
  \item{{\em No Sibling's Inner.} For each internal node, the computation of its outer code does {\em not} incorporate its sibling's inner code.}
  \item{{\em Inner-Outer Concatenation.} For each internal node, the computation of its outer code is based on the concatenation of both inner and outer codes of its parent node; {\em not} just outer.}
  \item{{\em No Inner for Leaves.} At a leaf node, only its outer, but {\em not} inner, code is used to decode an OBB.}
  \item{{\em No VQ-VAE.} No substructure prior.}
  \item{{\em No Hierarchy Resampling.} The initial hierarchy is used for all iterations.}
\end{itemize}
Vanilla GRASS~\cite{li2017} is obtained by omitting the VQ-VAE (substructure prior), omitting the use of inner codes when computing outer codes (input prior), and inferring decoding hierarchies from root codes alone (instead of reversing encoding hierarchies). This version did not converge in training to denoise noisy part assemblies, so we compare to only the (more powerful) variants above.

Over two relatively challenging categories, Chair and Bicycle, we plot merge error over increasing number of iterations, for all alternatives. (The plots for other categories are similar.) As can be observed in the plots, our full method achieves the fastest convergence and produces the lowest error for both datasets, compared to these alternatives. Specifically, we found that the incorporation of inner codes for leaf nodes, the VQ-VAE substructure prior, and hierarchy resampling are the most critical design choices for our method, since removing them causes the most significant performance drops (error increase). Since the curves for ``Inner-Outer Concatenation'' are quite close to those of our presented method, they are not plotted for clarity. However, we found training the former takes longer to converge.

To test how well we can merge substructures from {\em different} source shapes, we first show visual examples of challenging merges successfully performed by our method (Figure \ref{fig:hetero_visual}, also see supplementary). Our method can also merge parts from more than two source shapes, as shown in Figure \ref{fig:multi_source}.
Note that since the parts originate from different shapes, simple superimposition gives poor results and hence we do not need to add further noise. Next, we show how human raters compared our results in a preliminary user study.

\mypara{User study.}
We design two user studies to evaluate our shape composition algorithm in generating plausible shapes. It should be noted that a shape composition algorithm is but one part of the puzzle of modeling good shapes. Selecting {\em which} parts should be merged is another essential factor. Therefore, we perform two different studies. The first study is a sanity check on merging parts from a single complete shape. In the second, input parts are chosen from two or more shapes, to see if our method can merge parts from possibly incompatible structures. Each study surveyed $50$ participants (graduate students from various majors and backgrounds).

\begin{figure}[t!]
	\includegraphics[width=\linewidth]{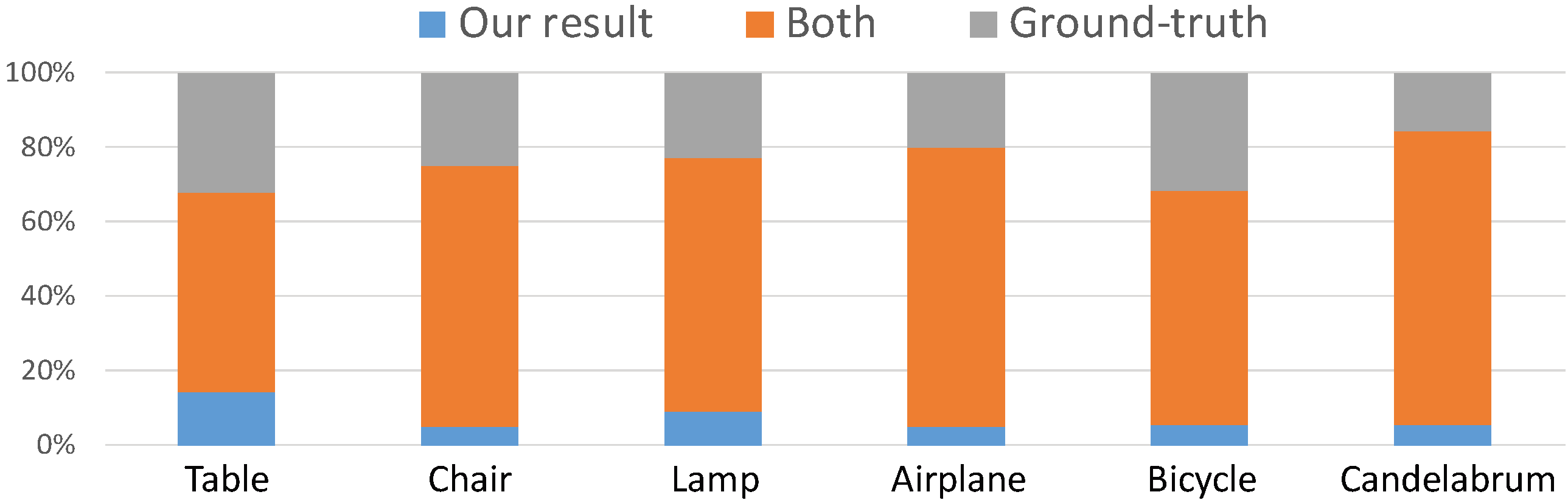}
	\vspace{-7mm}
	\caption{Human evaluation of shape plausibility between our merging results and ground truth.}
	\label{fig:userstudy}
	\vspace{-2mm}
\end{figure}

In the first study, we split one single shape into two groups of parts. The groups may overlap or omit some parts, and each part is randomly oriented and posed. We show human evaluators two shapes in random order: the result of merging the groups with our algorithm, and the original source shape (the optimal merging result). Note that the test shapes used in this study were \emph{not} included in the training set. We asked the evaluator: {\em ``which shape looks more plausible?''}, explaining that plausibility should be judged in the context of the part's category. In Figure~\ref{fig:userstudy}, we plot the preference percentages for $180$ different combinations of parts from $6$ shape categories. The results show that the merging artifacts produced by our method are barely recognizable by a human observer.

\begin{table}[t!]
  \begin{tabular}{c|c|c|c}
    {\bf Category} & {\bf \SCORES} & {\bf No VQ} & {\bf No Resampling} \\
    \hline
    \hline
    Airplane & {\bf 4.7/4.4} & 3.1/3.5 & 4.1/3.7 \\
    Bicycle & {\bf 4.2/4.1} & 3.3/3.0 & 4.0/3.8  \\
    Candelabrum & {\bf 4.7/4.6} & 3.9/3.7 & 4.3/4.0  \\
    Chair & {\bf 4.4/4.3} & 3.2/3.0 & 3.0/2.9  \\
    Lamp & {\bf 4.6/4.5} & 3.9/3.6 & 3.0/3.5  \\
    Table & {\bf 4.4/4.1} & 4.0/4.1 & 3.7/3.9
  \end{tabular}
  \caption{Human evaluation of heterogeneous structure merging statistics, as rated by human evaluators, on a scale ranging from 1 (very poor) to 5 (very good). Statistics for output plausibility ($P$) and merge quality ($M$) are presented separately as $P/M$.}
  \label{tab:hetero_stats_by_category}
  \vspace{-7mm}
\end{table}

\begin{table}[t!]
  \begin{tabular}{c|c|c|c}
    {\bf Condition} & {\bf \SCORES} & {\bf No VQ} & {\bf No Resampling}  \\
    \hline
    \hline
   {Single Sub-category} & {\bf 4.7/4.5} & 3.7/3.9 & 4.1/3.9  \\
   {Across Sub-categories} & {\bf 4.1/4.3} & 2.9/3.3 & 3.0/3.5
  \end{tabular}
  \caption{Human evaluation of heterogeneous structure merging statistics, within and across fine-grained sub-categories of chairs (e.g. swivel chairs, folding chairs, etc.). Statistics for output plausibility ($P$) and merge quality ($M$) are presented separately as $P/M$.}
  \label{tab:hetero_stats_by_disparity}
  \vspace{-7mm}
\end{table}

In the second study, each participant is first presented with two randomly chosen groups of parts from two different shapes, and then the merged result with a randomly chosen algorithm, all rendered as in Figure \ref{fig:hetero_visual}.
Since the two groups of parts may come from shapes with very incompatible structures, we do not set functionality or aesthetics as our goal, but focus only on merge quality. The participant rates the result from 1 (very poor) to 5 (very good). The rating targets two aspects:
(1)~{\em Plausibility}: how realistic the result looks independent of sources; and
(2)~{\em Merge quality}: how good the result is as a fusion of the source substructures.
Table \ref{tab:hetero_stats_by_category} reports results for all six categories, showing that \SCORES fares better in both plausibility and merge quality. In Table \ref{tab:hetero_stats_by_disparity}, we show results for the Chair set, with a split for intra- and inter- sub-categories of chairs (e.g. swivel chairs, folding chairs, etc). \SCORES achieves satisfactory quality in both cases, demonstrating the capability of our method in merging incompatible structures. Interestingly, VQ-VAE turns out to be more important than resampling for merging substructures from different shapes, in contrast to handling substructures from single shapes. This shows the importance of substructure snapping in latent VQ space for the adjustment of incompatible structures.

\begin{figure}[t!]
  \includegraphics[width=\linewidth]{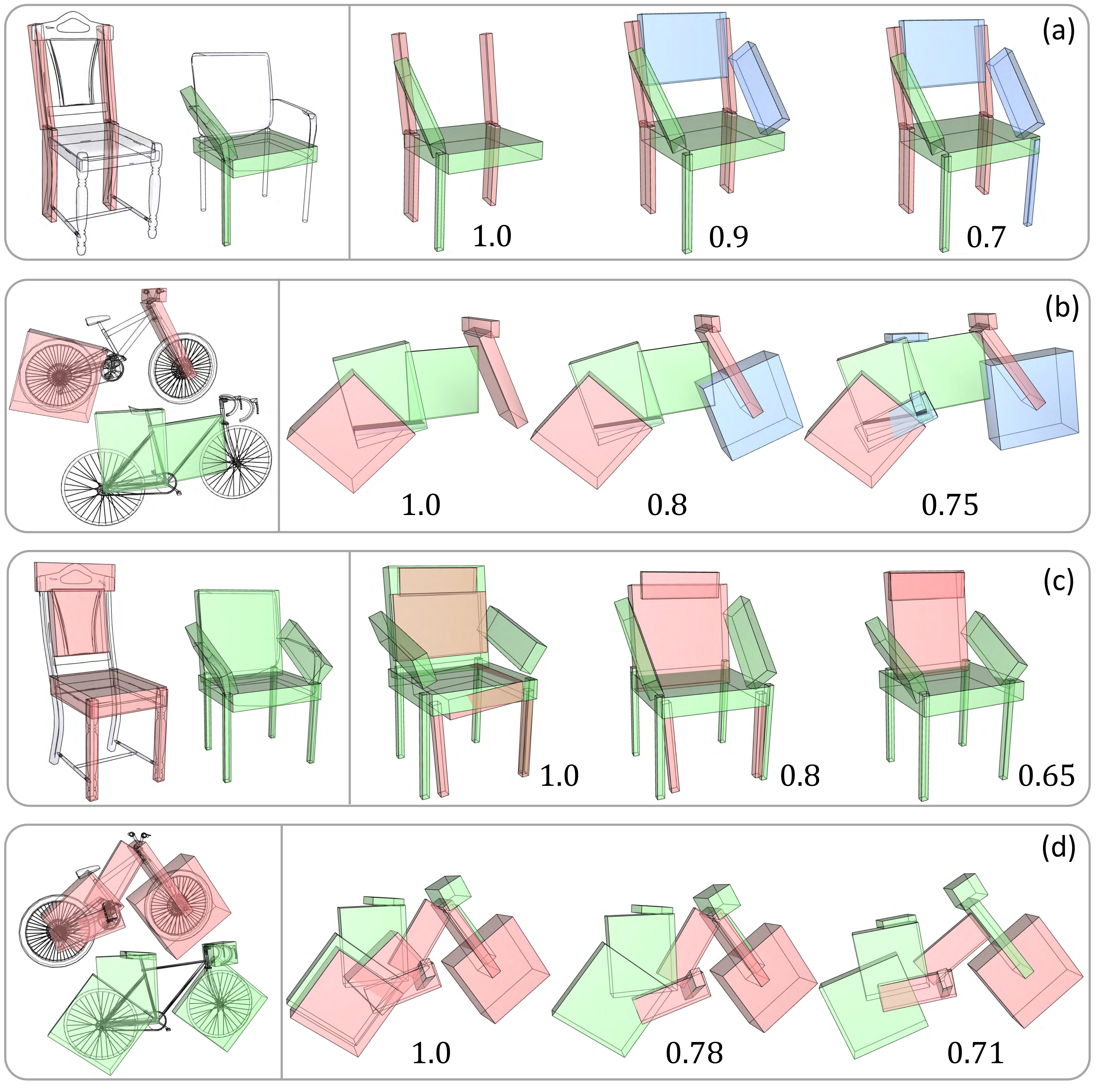}
  \vspace{-5mm}
  \caption{Examples of completing missing parts (a and b) and removing redundant parts (c and d), with different settings of $\eta_\text{T}$. $\eta_\text{T}=1$ corresponds to exact composition, while lower values lead to greater structure re-synthesis.}
  \label{fig:generative}
  \vspace{-3mm}
\end{figure}

\mypara{Flexible composition (with structure synthesis).} \SCORES differs from all prior structure merging methods in that it can add or remove parts to increase the plausibility of the resulting shape. Here, we demonstrate that our model is able to both complete missing parts and remove redundant (overlapping) parts in the input. In Figures \ref{fig:teaser} (left) and \ref{fig:hetero_visual} (d, h), we show merges with missing parts that are corrected by our method. In contrast, purely assembly-oriented methods, such as Fit \& Diverse \cite{xu2012}, cannot handle such corrections. Similarly, we also show examples (Figures \ref{fig:teaser} (right) and \ref{fig:hetero_visual} (c, g)) where the two merged groups have common (semantically similar) or overlapping parts, but these conflicts are resolved in the output. Figure~\ref{fig:generative} shows how our method can trade off between exact and flexible composition, controlled by the threshold $\eta_\text{T}$ determining when to invoke synthesis. As in the examples, a default $\eta_\text{T}$ of $0.7$ corrects most noticeable missing or redundant parts.
%

\begin{figure}[t!]
  \includegraphics[width=\linewidth]{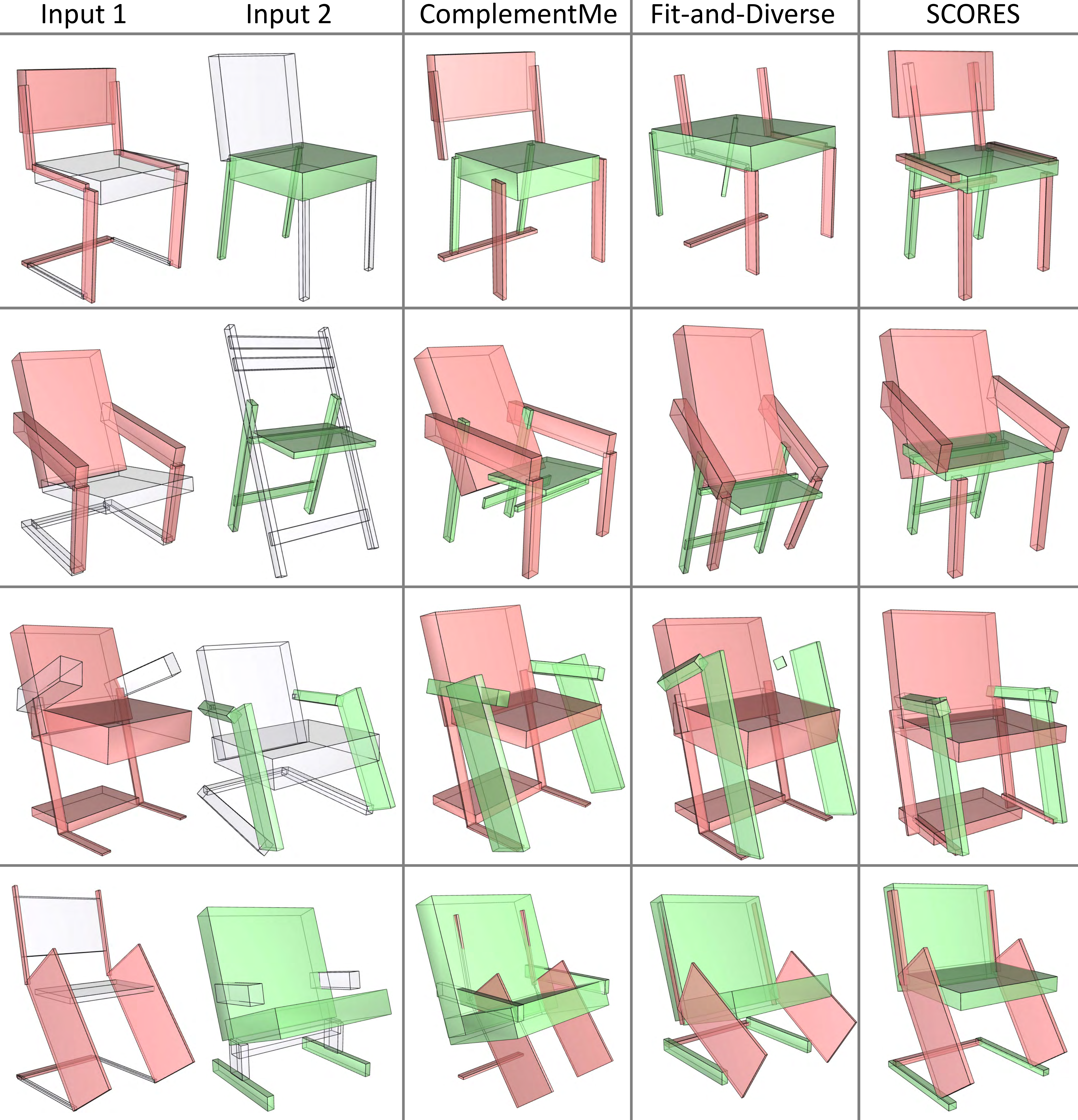}
  \vspace{-8mm}
  \caption{Visual comparison of shape composition results with ComplementMe~\cite{sung2017} and Fit-and-Diverse~\cite{xu2012}.}
  \label{fig:compare}
  \vspace{-3mm}
\end{figure}

\mypara{Comparisons.}
We compare our method with two state-of-the-art methods for shape composition: ComplementMe~\cite{sung2017} and Fit-and-Diverse~\cite{xu2012}. First, in Figure \ref{fig:compare}, we show a few representative visual comparisons. ComplementMe inserts new parts, {\em one at a time\/}, into a partial part assembly. It learns both a {\em part suggestion network\/} and a {\em part placement network\/} from training data, where the latter predicts the most likely position of a new suggested part based on the current assembly. We only compare our method with the placement network of ComplementMe. The parts to be inserted or composed are given in the two input shapes. Since ComplementMe can insert parts from either input shape into the other, we tried both directions and picked the better result obtained to shown in Figure \ref{fig:compare}. The less than satisfactory results by ComplementMe can be mainly attributed to its paradigm of {\em greedily\/} inserting one part at a time. The part placement network cannot adjust the current shape parts to better accommodate a new part. The placement is based on local criteria and not guided by an overall plausibility prior of the target shape.
%
%
In contrast, \SCORES jointly and hierarchically optimizes placements of all parts from the inputs, based on learned substructure priors and a plausibility loss, leading to better overall realism of the final 3D shape.

The Fit-and-Diverse tool of Xu et al.~\shortcite{xu2012} evolves a 3D shape collection to fit user preferences. The fundamental modeling operator is shape {\em crossover}, inspired by genetic algorithms, whereby corresponding parts from two shapes in the same category can be exchanged.
Crossovers are performed between shapes based on a fuzzy one-to-one part correspondence.
For shapes with incompatible structures, this correspondence can be quite unreliable, resulting in missing parts in the crossover results: see Figure~\ref{fig:compare}, first row. Hence, Fit-and-Diverse cannot handle high structural complexity and incompatibility. In contrast, \SCORES is correspondence-free and relies on a data-driven plausibility prior for structure optimization.

To quantitatively compare merging results between Fit-and-Diverse and \SCORES, we employ the structural plausibility measure of Zhu et al.~\shortcite{zhu2017}, based on multi-view depth images. This measure is learned by a deep neural network and trained on ShapeNet to predict the plausibility of a 3D shape with respect to an object category, i.e., ``how much the shape looks like a chair from multiple views".
We evolve the same input set using the Fit-and-Diverse framework, where the only difference is the crossover operator: either \SCORES, or the original scheme~\cite{xu2012}. We compute plausibility scores for the crossover results obtained by the two options, and repeat the experiment over 10 input sets from 3 categories: chair, airplane, and bicycle.
The overall result shows that the average plausibility score by \SCORES is about $34\%$ higher than that by Fit-and-Diverse. Scores for individual runs are in supplementary material.


We also make a comparison based on subjective measures of plausibility. We asked 50 human raters to judge the evolved shapes as plausible or not. The average percentage of plausible crossover outputs is $81\%$ for \SCORES and $70\%$ for Fit-and-Diverse.


\vspace{1mm}

Note that to ensure fairness of the above comparisons, our least-squares post-process for connecting docking points was omitted, so that only the arrangement priors were being compared.

\vspace{-1.5mm}
\mypara{Timing.}
The runtime performance of our shape composition depends on the size of the input part structures. For most examples, where the input parts number less than 30, our method typically converges within 20 iterations, in less than $1$ minute. About $75\%$ of the time is spent on hierarchy inference. Training takes 8 hours for VQ-VAE, and 12 hours for DAE, on an NVIDIA GTX 1080Ti.

\section{Applications}
\label{sec:apps}

\SCORES can be directly used for interactive assembly-based modeling: we show several example sessions with such an interface at the end of our accompanying video. In addition, we present two prototype applications demonstrating the applicability of \SCORES beyond interactive modeling: crossover for shape evolution and data-driven scan \mbox{reconstruction}.

\begin{figure}[t!]
  \includegraphics[width=0.99\linewidth]{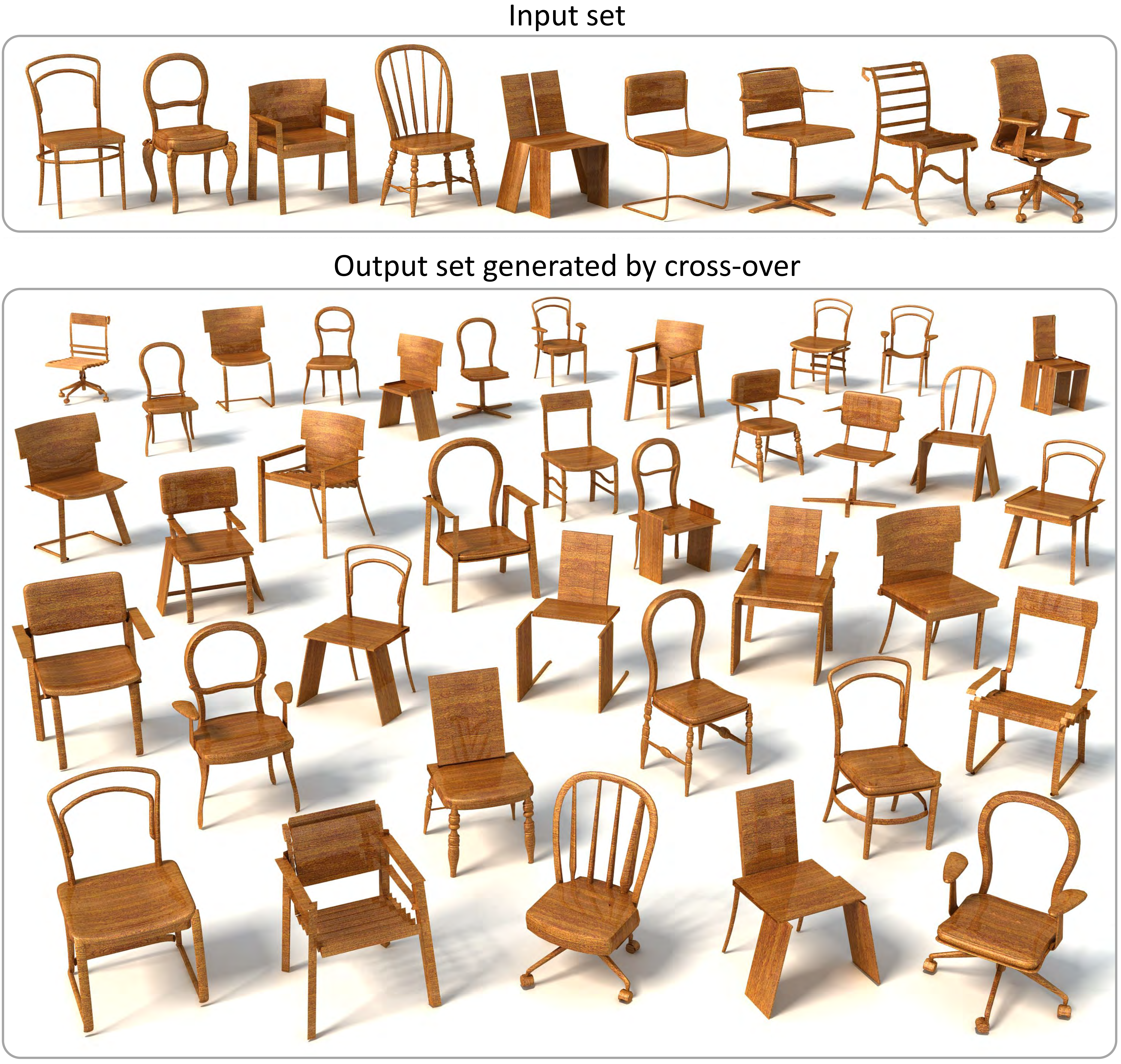}
  \vspace{-3mm}
  \caption{A chair collection (bottom) evolved from an input set of 9 chairs (top) using the Fit-and-Diverse framework of Xu et al.~\shortcite{xu2012}, where the crossover operations were performed by \SCORES.}
  \label{fig:crossover}
\end{figure}

\mypara{Set evolution.}
We can embed \SCORES into the Fit-and-Diverse~\cite{xu2012} modeling framework to improve the quality and inclusivity of shape crossover. Specifically, we use their fuzzy correspondence scheme to find corresponding parts in two shapes and swap the parts to initiate two shape composition tasks. We then apply \SCORES to carry out the tasks.
Figure~\ref{fig:crossover} shows a collection of $36$ chairs generated from a set of $9$ input chairs in this way.



\begin{figure}[t!]
  \includegraphics[width=0.99\linewidth]{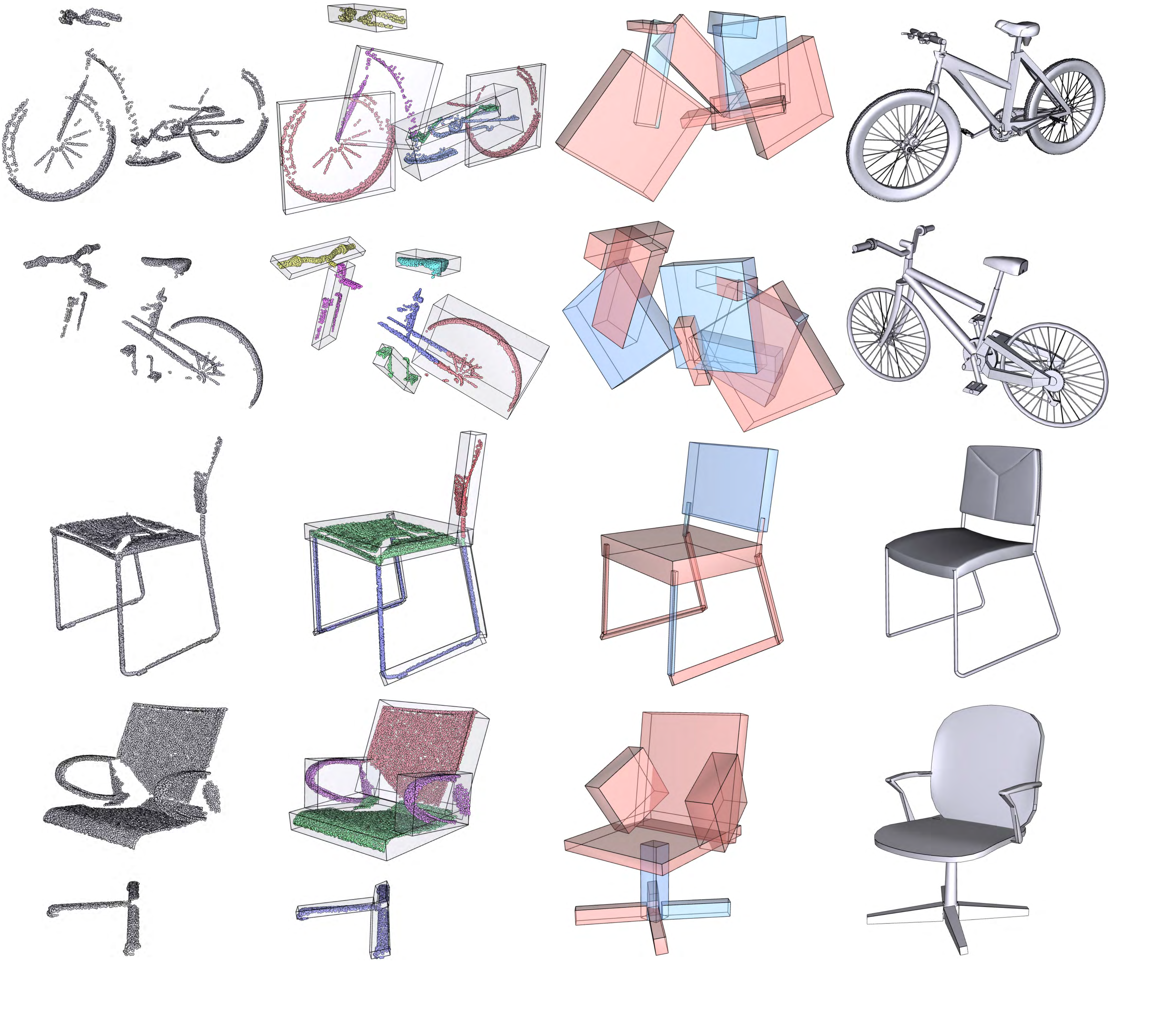}
  \vspace{-3mm}
  \caption{Complete shapes reconstructed from noisy, incomplete point clouds using \SCORES. 
Newly generated OBBs are marked in blue.}
  \label{fig:scan_completion}
  \vspace{-2.5mm}
\end{figure}

\mypara{Scan reconstruction.} Reconstructing complete, plausible shapes from incomplete and noisy point scans is an important, but ill-posed problem. Hence, it is natural to use data-driven priors for the task~\cite{Shen2012,sung2015}. Our shape composition network \SCORES can also be utilized for scan completion. As shown in Figure \ref{fig:scan_completion}, with strong shape priors learned from training data, \SCORES can reconstruct fine-grained shape geometry even for scans with significant missing portions. To produce these results, we first segment the scan using the supervised point clustering of Sung et al.~\shortcite{sung2015} (this is also the first step of their data-driven shape completion method). Then we fit OBBs to the point clusters and apply \SCORES to the collection of OBBs via substructure rectification. \SCORES can synthesize new parts based on learned substructure priors to improve the shape's overall plausibility, while Sung et al. rely on pre-defined global shape templates to fit the incomplete point clouds. However, \SCORES only performs structure rectification: to output fine-grained geometry, database parts are retrieved via SARF codes and transformed to fit output OBB parameters. Hence, the completed output may not precisely match the fine-grained geometry of the scan. This limitation can be addressed in the future by querying the part database with fine-grained segment geometries.
\section{Discussion, limitation, and future work}
\label{sec:future}

\rz{Our work on learning to compose 3D shapes has two key ``take-home'' messages. First, shape composition is more general than part insertion (e.g.~ComplementMe~\cite{sung2017}), part exchange/crossover (e.g.~Fit\&Diverse~\cite{xu2012}), and part connection, which are modeling paradigms adopted by previous works. In our general setting, we compose sets of parts whose union may not be a complete shape (hence, new parts need to be added) and there may be incompatibilities between parts (duplicates, significant geometry/topology mismatches). Our goal is to learn from data, to overcome all these issues and produce a plausible composed structure.} Our paper realizes this promise with a recursive autoencoder network (\SCORES). Results show that when trained on structured shapes from ShapeNet, the network is able to learn to compose inputs from multiple sources that may be corrupted by gross misalignments and missing or redundant parts.

\rz{The second insight is that it is possible to generate new shape structures, which are unseen in training data, through our shape composition, as long as the {\em substructures\/} in the input can be rectified from the training data. We believe this is a transferable concept that can empower generative models for other problems.}

\vspace{-0.5mm}
\mypara{Unlabeled OBBs.}
\rz{Our structural analysis and composition operate entirely on {\em unlabeled\/} OBBs without explicit encoding of part semantics. Like GRASS~\cite{li2017}, \SCORES works with shape structures that not only depend on where the part BBs are but also how they are related to each other in the context of a shape. The basic premise is that the location of a part in a shape and its spatial context can well characterize what that part is, even without the precise geometry of the part itself and its meaning. It would be interesting to see whether unlabeled OBBs are sufficient when we extend the analysis to 3D scenes, where shape parts would become objects in a scene, but the object-object relations in scenes appear to be a lot looser than part-part relations in 3D shapes.}

\mypara{Generative capabilities.}
Our network is not designed to be fully generative like GRASS~\cite{li2017}: it can synthesize novel structures but will only do so when part composition alone incurs too large a plausibility loss.
Specifically, when the VQ representation error is too large for an outer code, the ensuing structure synthesis is performed with a pre-trained denoising autoencoder, rather than based on VQ-VAE, as in GRASS. With VQ-VAE, we expect to obtain a more consistent solution, but we have found it to be hard to train. Part of the reason is that the outer code for an incomplete structure can be arbitrarily distant from that of its completed counterpart in the latent space. A possible remedy is to first learn a joint embedding of both complete and incomplete substructures, which we leave for future work.
Nevertheless, it is interesting to observe the trade-off between composition and synthesis; see Figure~\ref{fig:generative}. Like GRASS, \SCORES is trained with and operates on segmented 3D models, but we do not require part labels. Another limitation is that each input part is assumed to be complete and not partial. Relaxing this assumption may not be a must for assembly-based modeling, but could be desirable in other problem settings.

\mypara{Functionality, aesthetics, and creativity.}
These are all modeling criteria that our method does {\em not\/} explicitly account for. \SCORES learns substructure priors for shape composition; it is not trained to learn functionality, aesthetics, style, or creativity as there is no such knowledge embedded in the training data. Moreover, to obtain a plausible composite shape, a prerequisite is that the input parts could possibly lead to such a final shape. Currently, our method does not verify this; it simply minimizes the plausibility loss based on what \SCORES was able to learn from the data.

\begin{figure}[t!]
  \includegraphics[width=0.99\linewidth]{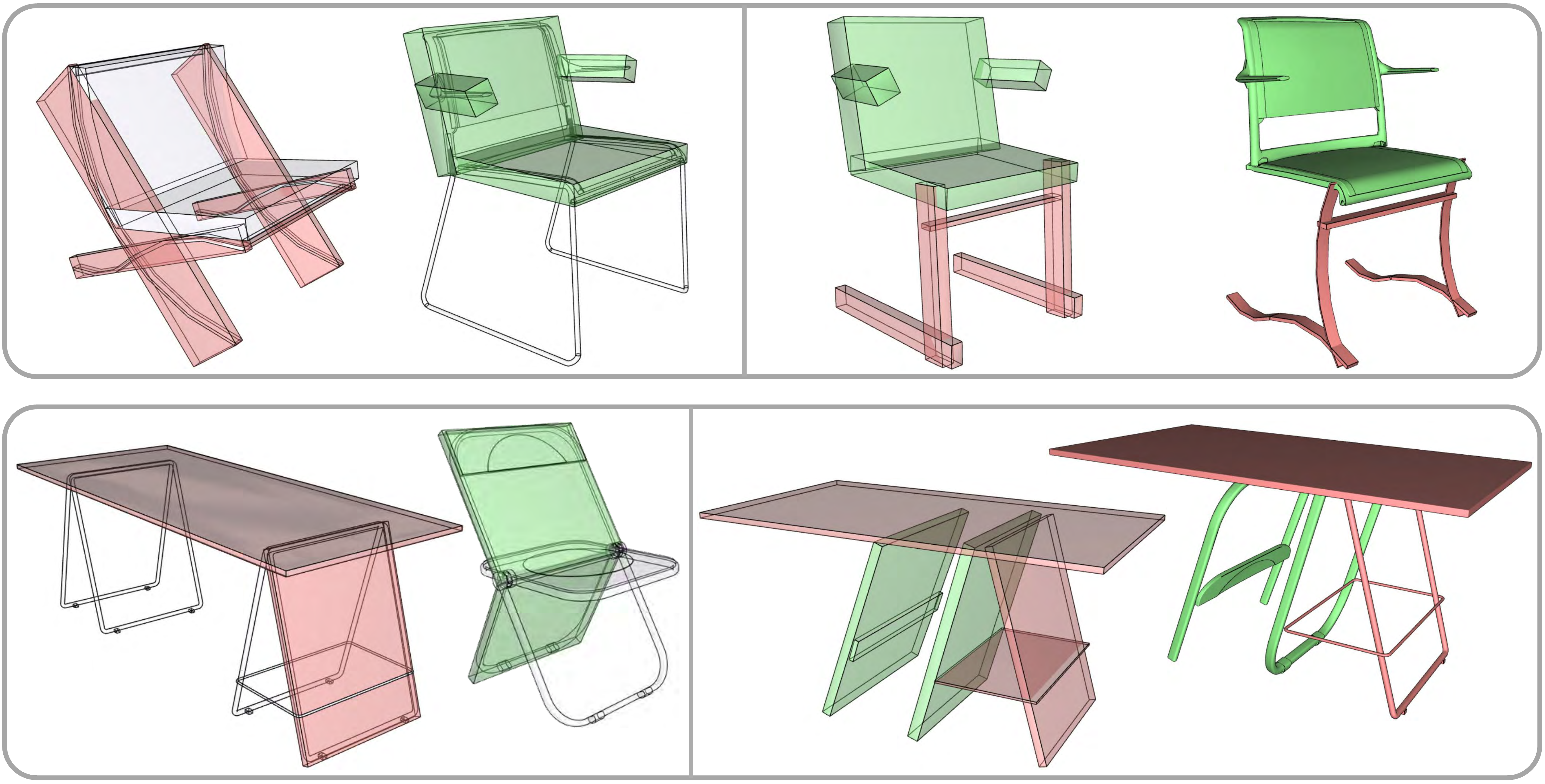}
\caption{Two typical failure cases. Top: As the training set contains too few folding chairs, our method tends to adjust the folding leg parts into a known, but undesirable substructure. Bottom: When input parts from chair and table (Figure~\ref{fig:hetero_visual}(j)) are assembled based on latent space learned for tables, the output is less plausible --- chair back cannot be placed well on a table.}
  \label{fig:failure}
\end{figure}

\mypara{Failure cases.}
Clearly, the success of our learning-based approach hinges on data. Figure~\ref{fig:failure} shows two typical failure cases of our method, which relies strongly on the discrete latent space learned from valid substructures. When an improper model is learned or utilized, implausible results may result. For example, when some substructure is absent or rare in the training set, our method tends to adjust it into a known but different substructure, without preserving the original structure. When performing cross-category part merging, it is possible that the substructure model, learned for one category, cannot accommodate parts from a different category, leading to implausible placement of those parts.

\mypara{Future work.}
Our work also opens the door to several other future directions. Rather than emphasizing composition over synthesis, we could reverse them and turn our network into a conditional generative model, allowing large portions of a 3D model to be synthesized based on sparse inputs. The inputs themselves can be expanded to include hand-drawn sketches or images. Generative image/sketch composition is also a promising avenue to explore.

\section*{Acknowledgements}
We thank the anonymous reviewers for their valuable comments and feedback.
This work is supported in part by grants from NSFC (61572507, 61532003, 61622212) and Hunan Province (2017JJ1002) for Kai Xu, NSERC Canada (611370) and Adobe gift funds for Hao Zhang, and China Scholarship Council for Chenyang Zhu and Renjiao Yi.

\bibliographystyle{ACM-Reference-Format}
\bibliography{shape_modeling}

\end{document}